\documentclass[12pt]{amsart}
\usepackage[
  hmarginratio={1:1},     
  heightrounded,
]{geometry}
\usepackage[foot]{amsaddr}
\usepackage{amssymb}

\usepackage{graphicx}
\usepackage{dcolumn}
\usepackage{bm}

\usepackage[utf8]{inputenc}
\usepackage[T1]{fontenc}
\usepackage{mathptmx}
\usepackage{etoolbox}

\usepackage{braket}
\usepackage{algorithm}
\usepackage{algpseudocode}

\usepackage[labelfont=bf]{caption}

\makeatletter
\def\@email#1#2{%
 \endgroup
 \patchcmd{\titleblock@produce}
  {\frontmatter@RRAPformat}
  {\frontmatter@RRAPformat{\produce@RRAP{*#1\href{mailto:#2}{#2}}}\frontmatter@RRAPformat}
  {}{}
}%
\algnewcommand{\LeftComment}[1]{\Statex \(\triangleright\) #1}
\makeatother

\title[ ]{Asynchronous Entanglement Routing for the Quantum Internet}

\author{Zebo Yang$^\dagger$, Ali Ghubaish$^\dagger$, Raj Jain$^\dagger$}
\address{$\dagger$ CSE, Washington University in St. Louis, St. Louis, MO, USA}
\email{zebo@wustl.edu, aghubaish@wustl.edu, jain@wustl.edu}

\author{Hassan Shapourian$^\ddagger$}
\address{$\ddagger$ Cisco Research, San Jose, USA}
\email{hshapour@cisco.com}

\author{Alireza Shabani$^\mathsection$}
\address{$\mathsection$ NSF Center for Quantum Networks, AZ, USA}
\email{alireza.shabani@gmail.com}

\date{\today}

\begin{document}
\begin{abstract}
With the emergence of the Quantum Internet, the need for advanced quantum networking techniques has significantly risen. Various models of quantum repeaters have been presented, each delineating a unique strategy to ensure quantum communication over long distances. We focus on repeaters that employ entanglement generation and swapping. This revolves around establishing remote end-to-end entanglement through repeaters, a concept we denote as the "quantum-native" repeaters (also called "first-generation" repeaters in some literature). The challenges in routing with quantum-native repeaters arise from probabilistic entanglement generation and restricted coherence time. Current approaches use synchronized time slots to search for entanglement-swapping paths, resulting in inefficiencies. Here, we propose a new set of asynchronous routing protocols for quantum networks by incorporating the idea of maintaining a dynamic topology in a distributed manner, which has been extensively studied in classical routing for lossy networks, such as using a destination-oriented directed acyclic graph (DODAG) or a spanning tree. The protocols update the entanglement-link topology asynchronously, identify optimal entanglement-swapping paths, and preserve unused direct-link entanglements. Our results indicate that asynchronous protocols achieve a larger upper bound with an appropriate setting and significantly higher entanglement rate than existing synchronous approaches, and the rate increases with coherence time, suggesting that it will have a much more profound impact on quantum networks as technology advances.
\end{abstract}

\footnote{This article has been accepted for publication in the AVS Quantum Science journal.}
\maketitle

\section{Introduction}

\begin{figure}
  \includegraphics[width=0.7\textwidth]{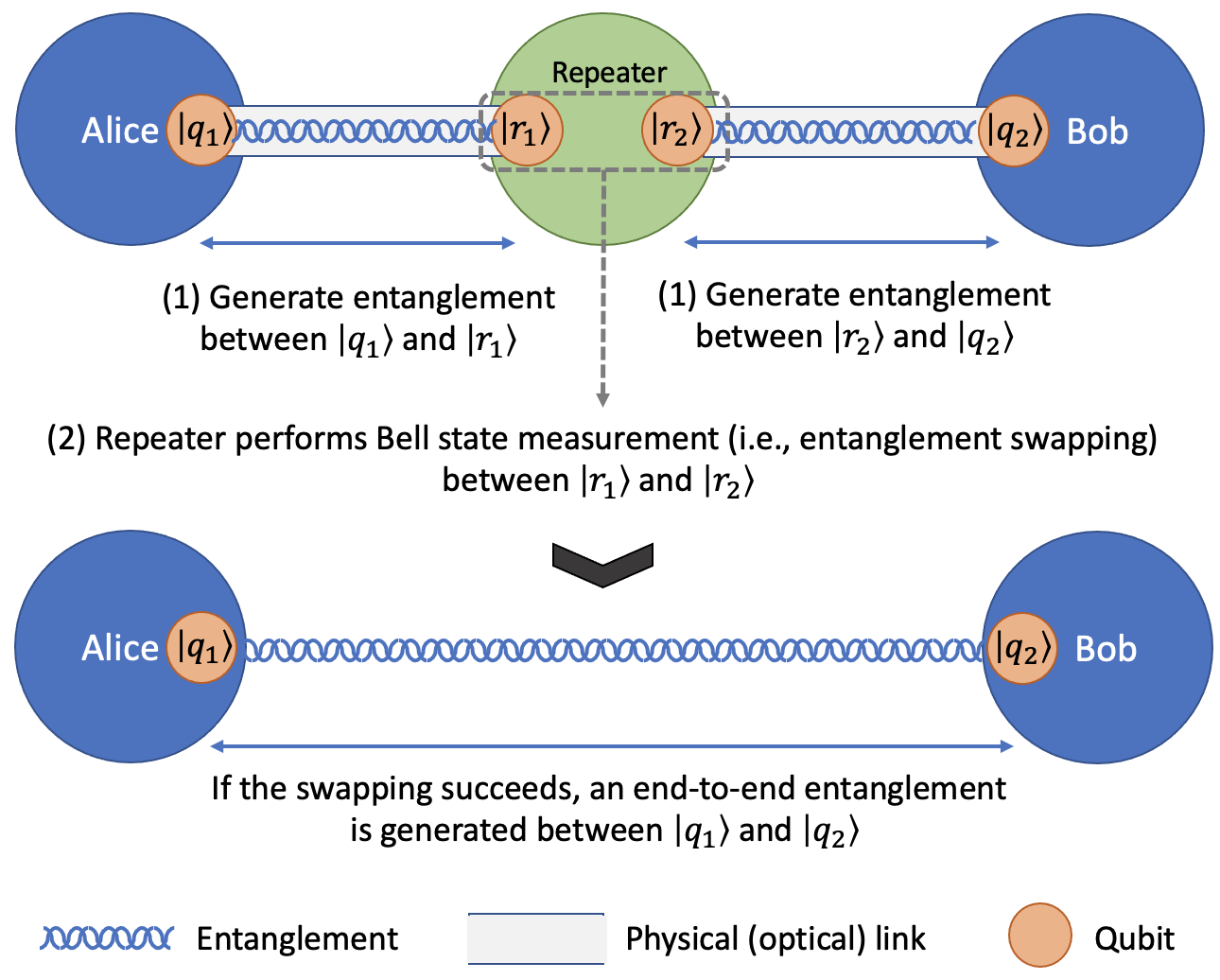}
  \captionsetup{width=1.0\linewidth}
  \caption{\textbf{Single-hop entanglement generation via entanglement swapping.} Alice and Bob intend to establish an end-to-end entanglement through a repeater positioned in between. The process involves two steps: \textbf{(1)} both parties create an entangled state with the repeater, resulting in two pairs of entangled qubits: Alice's $\ket{q_1}$ paired with the repeater's $\ket{r_1}$ and Bob's $\ket{q_2}$ paired with the repeater's $\ket{r_2}$. \textbf{(2)} the repeater conducts entanglement swapping, which essentially performs the Bell state measurement on $\ket{r_1}$ and $\ket{r_2}$ (see Subsection \ref{subsection:swapping}) assisted by classical communication, producing an end-to-end entanglement between Alice's $\ket{q_1}$ and Bob's $\ket{q_2}$.}
  \label{fig:entgmswap}
\end{figure}

Quantum networks \cite{kimble_quantum_2008, wehner_quantum_2018, yang_survey_2023} offer many possibilities for applications that go beyond what classical networks can accomplish. Among these are quantum key distribution (QKD) \cite{ekert_quantum_1991, lo_secure_2014}, sensing \cite{degen_quantum_2017}, secure communication \cite{gisin_quantum_2007}, clock synchronization \cite{komar_quantum_2014}, and distributed quantum computing \cite{beals_efficient_2013, van_meter_path_2016}. Quantum repeaters have been developed using various techniques to facilitate long-distance communication for these applications. These methods can be broadly categorized into two groups: the first involves transmitting encoded quantum information \cite{azuma_all-photonic_2015, niu_all-photonic_2022} and applying quantum error correction \cite{lidar_quantum_2013}, similar to classical repeaters. The second, hereafter referred to as the quantum-native repeaters (or in some literature, first-generation repeaters \cite{muralidharan_optimal_2016}), distributes long-distance entanglement \cite{briegel_quantum_1998, muralidharan_optimal_2016} through entanglement swapping \cite{pan_experimental_1998} (Fig. \ref{fig:entgmswap}), allowing end-users to carry out desired operations such as key distributions or secure communications. While a direct application of classical networking technologies to quantum networks by transmitting encoded quantum states may seem intuitive, the quantum-native repeater provides a more immediate solution to creating a network that utilizes end-to-end entanglement, as the former approaches necessitate a considerably impractical quantity of qubits for error correction purposes \cite{terhal_quantum_2015, preskill_quantum_2018}.

The development of direct-link entanglement generation has brought quantum networks closer to realization, with recent experiments demonstrating deterministic entanglement delivery between solid-state memories located in different physical locations \cite{humphreys_deterministic_2018}, facilitated by the use of telecommunication wavelengths to extend the distance between the locations \cite{tchebotareva_entanglement_2019, dreau_quantum_2018}. However, the rate of direct-link entanglement generation still decreases significantly over long distances due to photon loss and quantum state and operation instability resulting from unavoidable interactions with the environment. Quantum repeaters based on entanglement swapping are thus necessary to extend the entanglement generation (distribution) distance. However, they may not always be capable of achieving the desired fidelity levels required for certain applications \cite{wengerowsky_passively_2020, abruzzo_quantum_2013}. Entanglement purification (see Subsection \ref{subsection:purif}) can improve fidelity, but it has a trade-off. It reduces the number of entangled pairs shared between neighboring nodes on the network and requires a substantial amount of classical communication overhead \cite{dur_quantum_1999}. Entanglement swapping and purification form a quantum-native repeater scheme analogous to the link layer of classical networks (see Appendix \ref{appe:networkstack}).

On top of the quantum-native repeater scheme, a network layer can be established to ensure an effective architecture \cite{acin_entanglement_2007, perseguers_quantum_2010} for quantum networks, thereby enabling the full network capabilities. The quantum network layer plays a crucial role in determining the routing principles, i.e., quantum-native routing, that facilitate requests for generating end-to-end entanglements between remote stations \cite{pant_routing_2019, li_effective_2021, patil_entanglement_2022}. Various metrics have been suggested to direct the routing process \cite{shi_concurrent_2020, caleffi_optimal_2017}, and the entanglement rate — the number of entanglements generated per unit time — is a prevalent metric akin to throughput in  classical networks, with entanglements (as the communication resources) replacing bits. In essence, designing a quantum-native routing protocol aims to determine the optimal path while maximizing the entanglement rate between any two nodes under specific constraints, such as operation probability, coherence time, and link capacity. It should be noted that a classical network is assumed to be present in the quantum network, responsible for performing routing computations and broadcasting routing information. In other words, each node in the quantum network must be able to communicate with one another via a classical network, c.f., local operations and classical communications (LOCC). This assumption can be safely made given the expectation of a hybrid quantum-classical network and is a common practice in quantum routing techniques \cite{pant_routing_2019, li_effective_2021, patil_entanglement_2022, shi_concurrent_2020, caleffi_optimal_2017, hahn_quantum_2019}. After the paths have been established with the aid of LOCC, the intermediate nodes (repeaters) carry out entanglement swapping and purification. The resulting end-to-end entanglements connecting the endpoints (end users) are then used as the communication resources for any desired operations. The nature of end-to-end entanglement resembles connection-oriented circuit-switching, in contrast to the connection-less packet switching commonly used in classical networks, which was also proposed for quantum networks \cite{diadamo_packet_2022}. To avoid confusion, the term "quantum-native" is explicitly used to denote that the technique applies only to quantum networks (to distinguish from the repeater approaches that relay qubits) and is not independent of classical networks.

\begin{figure}
  \includegraphics[width=0.8\textwidth]{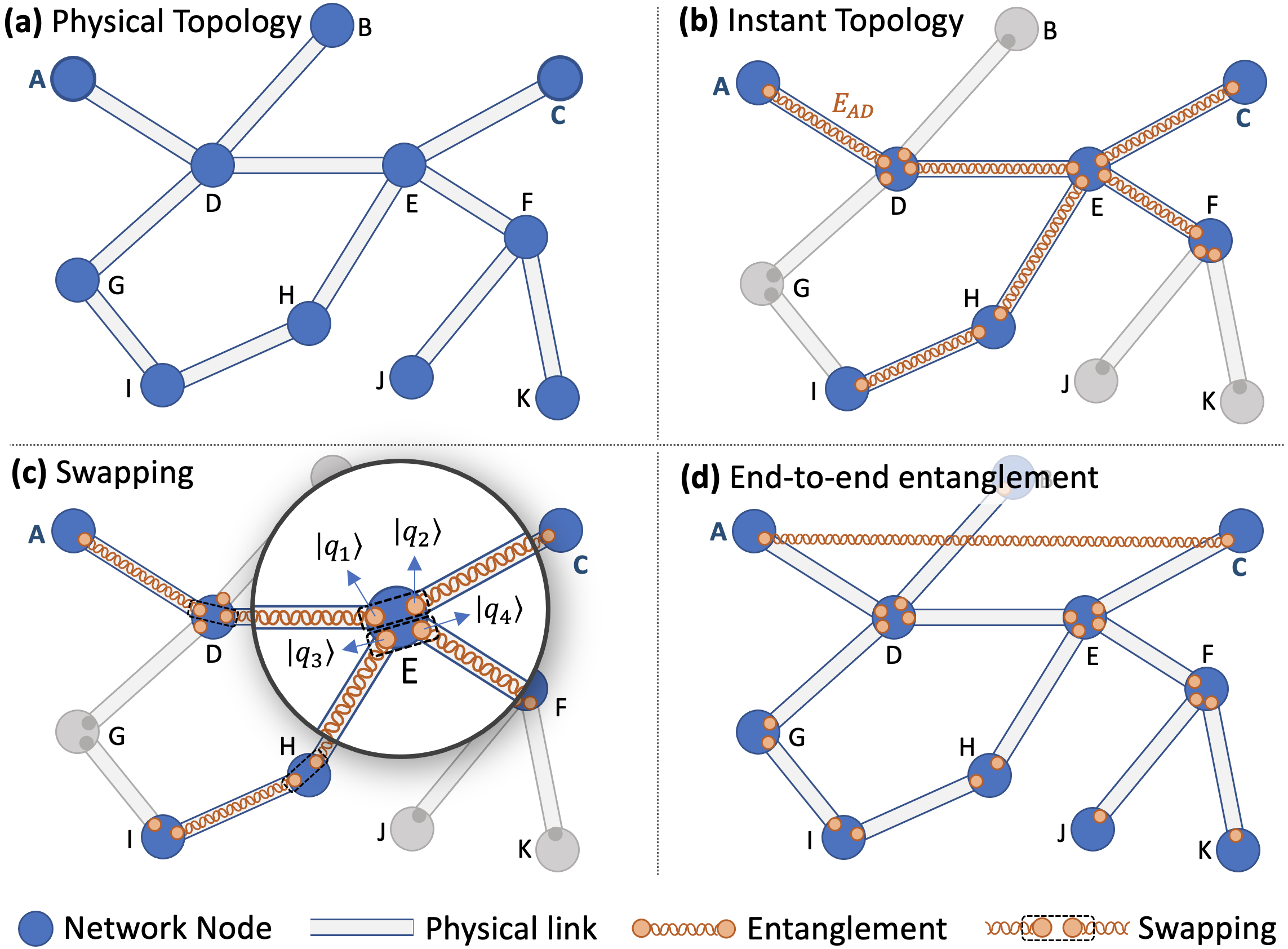}
  \captionsetup{width=1.0\linewidth}
  \caption{\textbf{End-to-end entanglement generation.} Nodes A and C in the network intend to establish an end-to-end entanglement. \textbf{(a)} The network consists of nodes connected by optical communication links, forming the \textit{physical topology}. \textbf{(b)} In the external phase, nodes connected by the same optical link attempt to generate direct-link entanglement, resulting in an \textit{instant topology} of entanglement links (the orange helical links). For example, Nodes A and D are connected by the entanglement link $E_{AD}$. \textbf{(c)} In the internal phase, all repeaters involved in two pairs of entanglements perform entanglement swapping on the qubits closer to the source or destination. Node E, for instance, has four qubits, with each qubit entangled with one of its four neighboring nodes ($E_{DE}$, $E_{HE}$, $E_{EF}$, and $E_{EC}$). It performs swapping on $\ket{q_1}$ and $\ket{q_2}$ since they are connected to the nodes closer to A and C, respectively. \textbf{(d)} By consuming all entanglements in the instant topology, A and C successfully obtain an end-to-end entanglement. Since each node is only aware of itself and its neighbor's entanglement-link statuses, Node E does not know if the path A-D-E-C would succeed. It also performs swapping on $\ket{q_3}$ and $\ket{q_4}$ to increase the success rate of achieving an end-to-end entanglement.}
  \label{fig:topology}
\end{figure}

The routing process in existing approaches \cite{pant_routing_2019, li_effective_2021, patil_entanglement_2022, shi_concurrent_2020} is divided into two phases, namely external and internal, each assigned a specific time slot. During the external phase, pairs of network nodes with a direct connection try to establish entanglement using a physical link, resulting in an "instant topology" (Fig. \ref{fig:topology}a and \ref{fig:topology}b) consisting of direct-link entanglements. In the internal phase, assuming that nodes are only aware of the direct-link entanglement generation results of their neighbors (see Subsection \ref{subsection:globalocal}), each repeater node swaps entanglement blindly to reach an end-to-end entanglement between the two endpoints that request to communicate (Fig. \ref{fig:topology}c and \ref{fig:topology}d). If the end-to-end entanglement is not achieved, the two phases are repeated. It is essential to synchronize both phases to ensure that entanglement "links" remain available for the internal phase. However, this synchronization and blind attempts consume all entanglements in each time slot, leading to a low entanglement rate. Furthermore, there are no obvious solutions for connecting networks of networks in these existing approaches.

In this paper, we propose a set of asynchronous routing protocols that can find the path between any two endpoints in a network with a much higher entanglement rate than existing approaches. Also, the entanglement rate increases with coherence time, and there are gateways in the network to connect networks of networks. It removes the need for synchronized executions by iterating an instant topology consisting of entanglement links. The instant topology can be in the form of a graph that can be maintained in a distributed manner, such as a destination-oriented directed acyclic graph (DODAG) [12] or a spanning tree. Such techniques have been well-studied for classical lossy networks, which share the similarity of losing a connection (in a quantum context, an entanglement in the instant topology) with a probability depending on the quality of the physical link. Our theoretical analyses and simulations demonstrate that our asynchronous protocols achieve a larger upper bound and significantly higher entanglement rate compared to existing synchronous approaches. The entanglement rate increases as coherence time increases. Additionally, a tree-like instant topology such as DODAG allows for the connection of networks of networks through the root nodes. These findings suggest that our proposed protocols will have a significant impact on the development of a Quantum Internet as technology advances (as coherence time increases and networks expand in scale).

The rest of the paper is organized as follows. First, we define the problem statement of this research by introducing the routing metric and the settings of the existing routing scheme we aim to optimize. Then, we propose the framework and two example algorithms for the asynchronous routing protocols that maintain an instant topology in a distributed fashion. Next, we evaluate the routing performance from various perspectives and provide a detailed explanation of a sample routing result with DODAG. Subsequently, we analyze the simulation results to examine the effectiveness of our scheme under different conditions and establish its superiority. Finally, we conclude the paper by discussing the limitations, the extent of usability, and potential areas for future research and expansion.

\section{Preliminaries}
This section provides an overview of key concepts and methodologies relevant to our research. It offers essential background to ensure readers grasp the subsequent discussions and the context of our research question.

\subsection{Entanglement swapping}
\label{subsection:swapping}
\begin{figure}
  \includegraphics[width=0.8\textwidth]{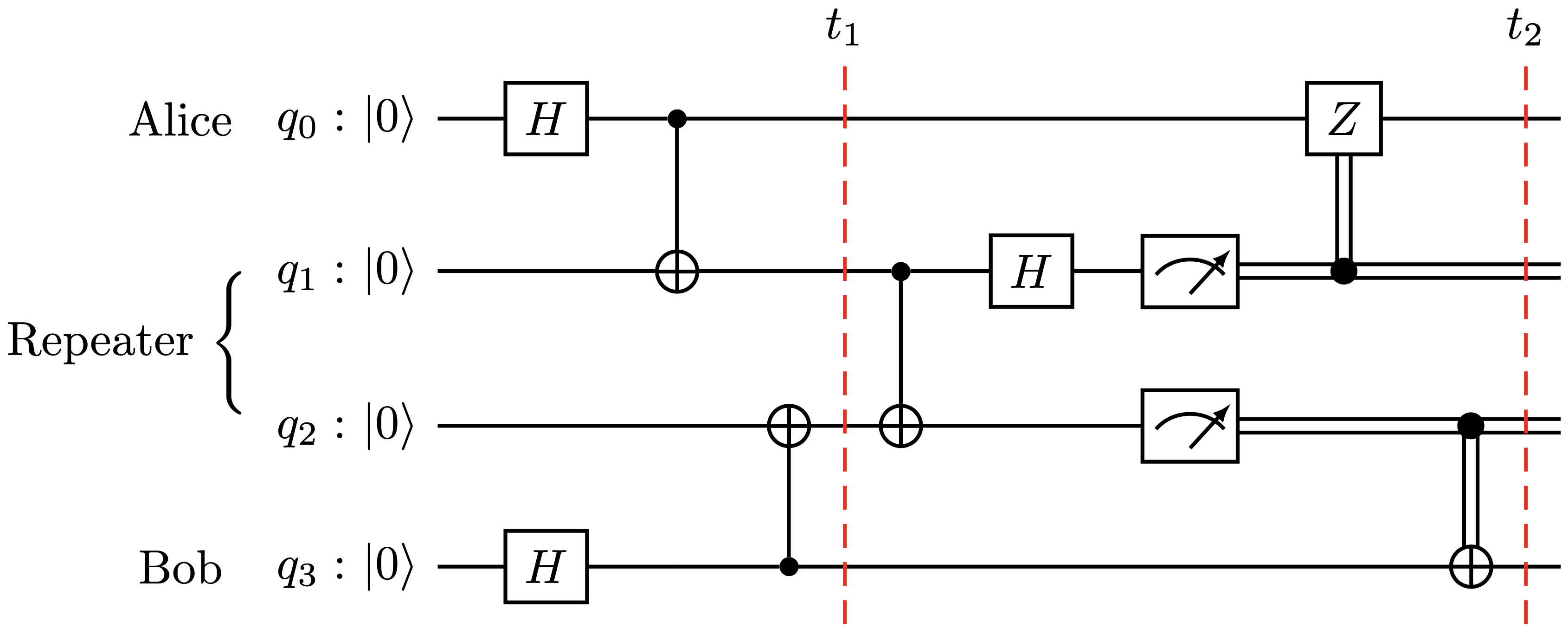}
  \caption{\textbf{The circuit of entanglement swapping.}}
  \label{fig:swapping_circuit}
\end{figure}
Entanglement swapping, in essence, is to perform a Bell state measurement on two pairs of entangled qubits. In the circuit depicted in Fig. \ref{fig:swapping_circuit}, three parties are involved: Alice, Bob, and a repeater. Alice possesses a qubit $q_0$, the repeater has qubits $q_1$ and $q_2$, and Bob has a qubit $q_3$. Before $t_1$, operations (i.e., Hardamard and controlled-not gates) are carried out to produce two pairs of entangled qubits: one between Alice and the repeater, $|q_0q_1\rangle$, and another between the repeater and Bob, $|q_2q_3\rangle$. Following this, the repeater executes a Bell state measurement, which is used initially to identify the specific Bell state that an entangled pair is in. After that, at $t_2$, the two entangled pairs $|q_0q_1\rangle$ and $|q_2q_3\rangle$ are swapped to form a new, single pair of entangled qubits, $|q_0q_3\rangle$, which connects Alice and Bob with an entangled pair (i.e., an entanglement link).

\subsection{Entanglement purification}
\label{subsection:purif}
\begin{figure}
  \includegraphics[width=0.8\textwidth]{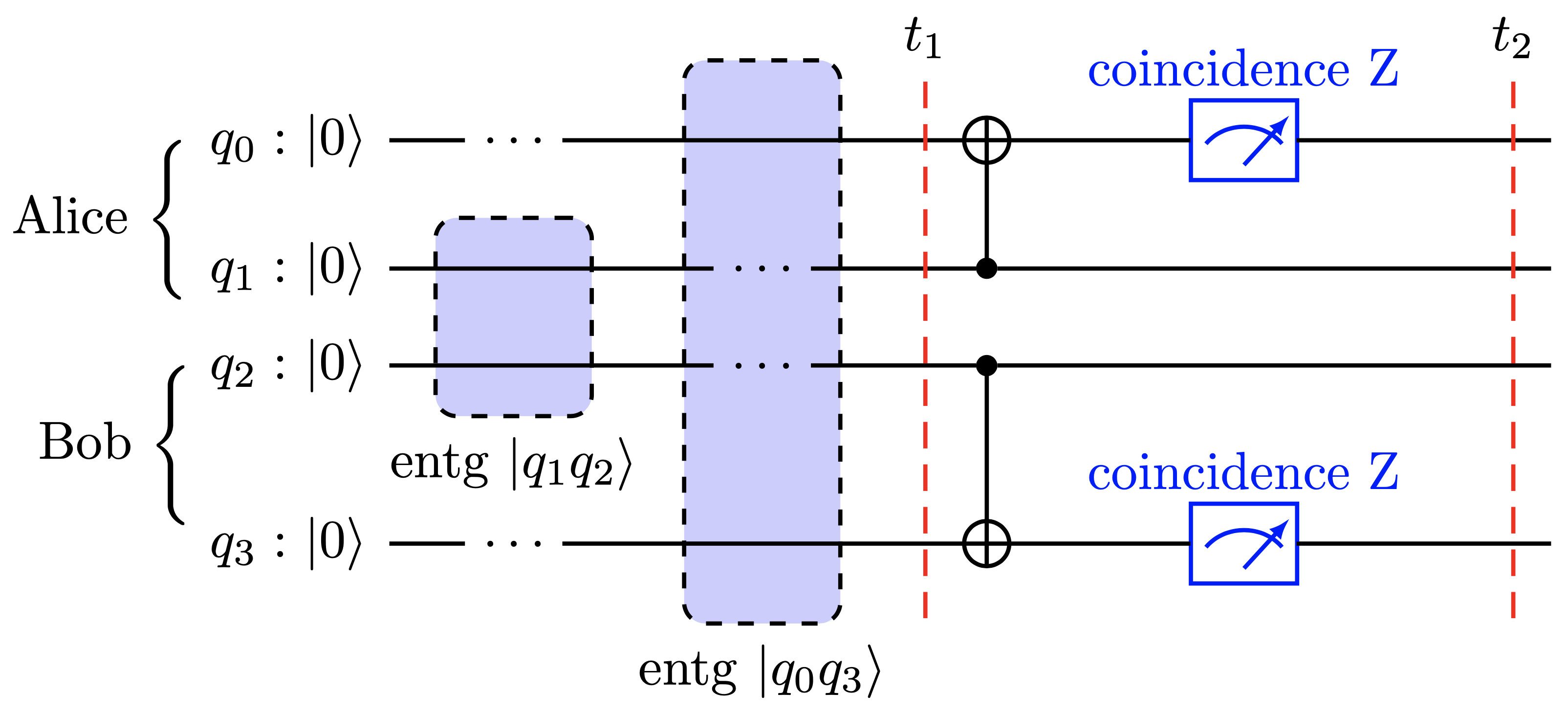}
  \caption{\textbf{An example purification circuit with link capacity two.}}
  \label{fig:purify_circuit}
\end{figure}
At time $t_1$, Fig. \ref{fig:purify_circuit} demonstrates that Alice and Bob possess two entangled pairs, namely $|q_1q_2\rangle$ and $|q_0q_3\rangle$, indicating a link capacity of two between them. This implies that two raw entanglements with moderate fidelity will be utilized to produce a single entanglement with higher fidelity. Alice owns $q_0$ and $q_1$, while Bob owns $q_2$ and $q_3$. Both $|q_1q_2\rangle$ and $|q_0q_3\rangle$ are considered low-quality (raw entanglements). Successful purification is determined by coincidence measurements in the Z basis. After the purification process, one of the entangled pairs is consumed while the other becomes a higher-quality entanglement. In this example, $|q_0q_3\rangle$ is consumed, and $|q_1q_2\rangle$ is preserved.

\subsection{Global and local knowledge of direct links}
\label{subsection:globalocal}
For a quantum routing protocol to be effective, it must be capable of identifying the optimal path as and when required. If a previously selected path becomes unavailable, the proposal must be able to identify an alternative path from the current location to the destination. Since the global knowledge of the entire network's entanglement links is often not up-to-date, the protocol must identify the best adjacent node based on its local knowledge of the entanglement links connected to it. Although the shortest path in an instant topology can be easily identified if a global view is available, it is unrealistic to assume every node has the latest version of it (e.g., to assume every node instantly knows the instant topology created by nodes ${A, D, E, H, I, F, C}$ in Fig. \ref{fig:topology}). The status of all direct-link entanglements can only be propagated through classical networks, which is relatively slow. In large networks, the instant topology may have already been altered by the time the global view has been broadcast to every network node. As a result, routing should be constrained to ensure that each node is only aware of the entanglement link states of its neighbors (e.g., node $D$ only knows the direct-link entanglement generation results of links $D-A$, $D-B$, $D-G$, and $D-E$. Specifically, $D-A$ and $D-E$ have been successful, while the other two have failed).

\subsection{Finding disjoint paths}
\label{subsection:disjointpaths}
In a network, disjoint paths refer to two or more paths between a pair of nodes (i.e., source and destination) that do not share any common node or link/edge except for the source and destination. Here, we consider link-disjoint paths where two paths do not have any link or edge in common.
Finding disjoint paths is necessary in reality due to the current low link capacity in quantum networks. Assuming global knowledge of the instant topology, finding even two disjoint shortest paths in many scenarios remains an NP-complete problem, such as the min-min (i.e., to minimize the length of the shorter path of the two disjoint paths), min-max (i.e., to minimize the length of the longer path of the two disjoint paths), and min-sum (i.e., to minimize the sum of the length of the two disjoint paths) problems. Besides single-cost min-sum, others are NP-complete. In addition, it is NP-hard to obtain a $\rho$-approximation to the optimal solution for any $\rho > 1$. More details can be found in Reference \cite{xu_complexity_2006}. This increases the time needed for classical operations to find the shortest paths. Calculating paths and propagating the global view of the instant topology within the entanglement coherence time is unrealistic. However, this problem will be eased in the future if there is a high link capacity.

\subsection{Link capacity and desired fidelity}
\label{subsection:linkfidelity}
\begin{figure}
  \includegraphics[width=0.8\textwidth]{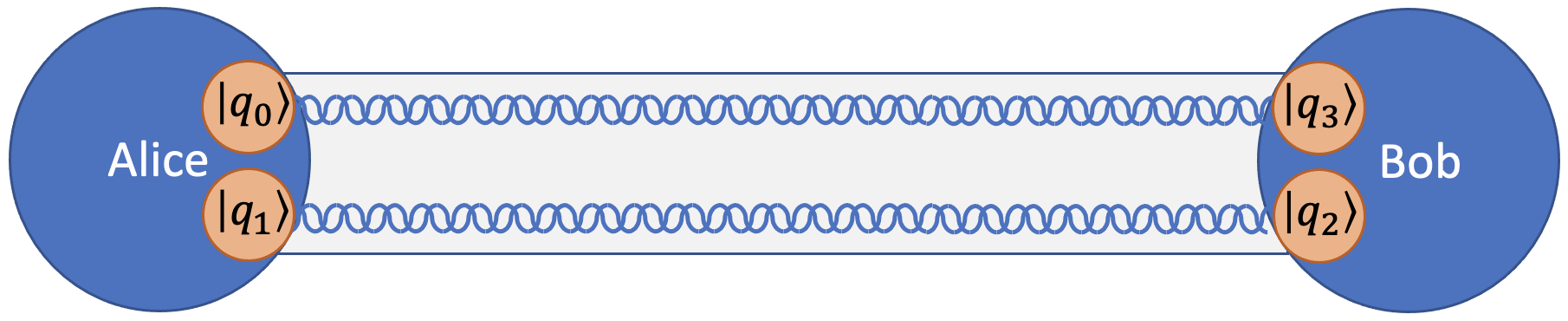}
  \caption{\textbf{A link between Alice and Bob with capacity two.}}
  \label{fig:link_capacity}
\end{figure}
The maximum number of entanglements that can be generated between two parties is referred to as link capacity. As illustrated in Fig. \ref{fig:link_capacity}, Alice and Bob each have two qubits, which means they can generate two entangled pairs, leading to resource/qubit allocation and scheduling issues. For more information, please refer to Reference \cite{li_effective_2021}. However, if there is redundant link capacity, a link in the instant topology can last longer since the other pairs of qubits (besides the one being used) will continually attempt entanglement generation, providing backup direct-link entanglements. Redundant entanglements in one edge can also be utilized to generate multiple paths and support simultaneous connection requests. However, having four qubits does not necessarily mean we can achieve the maximum of two entanglements since various entanglement-based applications, such as QKD, require a minimum fidelity level. Some entanglements must be used up to attain the desired fidelity, as explained in Subsection \ref{subsection:purif}. Consequently, the actual link capacity should be $C/E(f)$. Assuming homogeneous link and node quality, this is a constant factor applied to the entanglement rate. In the main text, without loss of generality, we set the link capacity $C = 1$ and the number of entanglements for purification $E(f) = 1$ since they are constant as average numbers. This simplifies the routing problem and allows us to focus on the entanglement rate change by maintaining a distributed graph.

\section{The Entanglement Routing Problem}
Quantum networks have several constraints that do not apply to classical networks. One such constraint is the probabilistic nature of entanglement generation and swapping performed by repeaters. Without loss of generality, we consider the generation of the maximally entangled states (Bell states) between adjacent nodes and the operation of Bell state measurement at repeater nodes to swap entanglement. Each instance of generation or swapping has a chance of succeeding or failing, leading to probabilistic and unpredictable routing paths. As pathfinding takes place and direct-link entanglements are generated, the instant topology changes constantly, emphasizing the need for an adaptive routing approach to creating an end-to-end entanglement. Finally, once an end-to-end entanglement is consumed, the network must establish a new one for a new connection request. In short, end-to-end entanglements allow the creation of dedicated unicast channels between node pairs, regardless of their position in the network. As mentioned, this characteristic of entanglement resembles connection-oriented circuit-switching rather than the connection-less packet switching used in classical networking. Here, we present a definition of the quantum-native routing problem.

Let us consider a graph $G(V, E)$ that represents the physical topology of the quantum network (Fig. \ref{fig:topology}a). In this graph, each node $v \in V$ represents a quantum-native repeater, while each edge $e \in E$ denotes an optical link that connects two repeater nodes. The instant topology (Fig. \ref{fig:topology}b) is denoted as $G'(V', E')$, which is a subgraph of $G$. $E'$ is the set of direct-link entanglements, and $V'$ are the nodes that are connected by the entanglements in $E'$. Additionally, each edge (link) $e$ can support multiple entanglement generations with multiple qubit pairs between adjacent nodes, where the link capacity $C(e)$ of link $e$ is defined as the maximum number of entangled pairs generated between adjacent nodes connected by $e$. Each node contains a finite number of qubits (depicted as orange dots in Fig. \ref{fig:topology}), denoted as node capacity $C(v) \ge \sum_{e \in N(v)} C(e)$ where $N(v)$ refers to the set of links connecting node $v$. To achieve a desired fidelity $f$, we need $E(f)$ entanglements for purification, and each entanglement has an average coherence time, $T_{co}$. For simplicity, we assume that the link capacity $C$ and purification number $E(f)$ are both equal to 1. This implies that, as depicted in Fig. \ref{fig:topology}, only one qubit is present on one side of the physical link. Nonetheless, varying link capacities and purification numbers will not impact the evaluation of the routing scheme as they simply introduce a constant multiplier to the entanglement rate. The entanglement rate will increase linearly relative to the actual link capacity, as detailed in Subsection \ref{subsection:linkfidelity}. Given our assumption that $C/E(f) = 1$, we can omit the $C/E(f)$ multiplier from each entanglement rate equation.

In addition, for each entanglement generation between a direct link $e$, there is a probability $p(e)$ that the generation will succeed, while each entanglement swapping at a repeater $v$ has a probability $q(v)$ of success. For convenience, we assume homogeneous entanglement generation probability, denoted by $p$, and entanglement swapping probability, denoted by $q$, on all network links and nodes. Finally, the entanglement rate, denoted by $\xi$, is the performance metric indicating the number of end-to-end entanglements (Bell states) that can be generated in a unit time $T$. $T$ must satisfy $T \le T_{co}$. Existing approaches usually consider the time required for both internal and external phases (as discussed in the Introduction Section) to be one unit time and assume it to be longer than the coherence time. To facilitate comparison, we set the unit time as $T = \frac{T_{co}}{n}$, where $n \in \mathbb{Z}$ and $n \ge 1$.

With these preliminaries, the problem of quantum-native routing is to design an effective algorithm that maximizes the entanglement rate under the constraint that nodes only possess local knowledge of adjacent links in the instant topology (For example, Node E in Fig. \ref{fig:topology}b only knows the direct-link generation results of $E_{ED}$, $E_{EH}$, $E_{EF}$, and $E_{EC}$). Note that it is unrealistic to assume that a global view of the instant topology is known, which may allow an algorithm to determine the shortest paths in the instant topology. In existing protocols such as \cite{pant_routing_2019, li_effective_2021} that depend on global knowledge of the instant topology, entanglements need to possess longer coherence times as the network size (i.e., the number of network nodes) expands. This is because direct-link entanglement information takes a relatively long time to traverse the entire network. Consequently, in a large-scale quantum network, direct-link entanglements may decay by the time all direct-link generation results are collected. Thus, a distributed approach is necessary to find optimal paths with local knowledge of the instant topology. 

Nonetheless, even if we assume to have a global view of the instant topology, it is still NP-complete to find multiple disjoint shortest paths (for robustness) in many scenarios (see Subsection \ref{subsection:disjointpaths}), given the restricted resources available at each link \cite{chakraborty_distributed_2019}. However, many existing approaches assume that the global view of the instant topology is known, such as \cite{li_effective_2021, patil_entanglement_2022, hahn_quantum_2019, shi_concurrent_2020}. A recent work by Pant et al. \cite{pant_routing_2019} proposes a blind search routing strategy with local knowledge of the instant topology, which, however, requires synchronized phases and consumes all direct-link entanglements in each attempt, as shown in Fig. \ref{fig:topology}c and \ref{fig:topology}d, resulting in a low entanglement rate. To ensure efficiency, entanglement swapping within a repeater should only occur when a path that includes that node is selected. Our approach, therefore, involves maintaining a distributed graph and reserving entanglements not used in the current paths for future use.

\section{Asynchronous Routing Scheme}
In essence, our method of asynchronous routing functions on an event-driven basis. It establishes and maintains the instant topology as a distributed graph (such as DODAG or spanning tree) and reacts to connection requests (events) by following the path determined in the distributed graph, such as tracing the path through the most current DODAG's root. The crux of the matter is the former service, which outlines how to preserve the instant topology as a distributed graph that determines a route for any given pair of nodes. In particular, we illustrate the protocol using two sample graphs: DODAG and spanning tree. Protocol \ref{alg:pf} presents the algorithmic steps of the routing processes, including quantum operations (QO), such as entanglement generation and swapping, and classical operations (CO), such as event listening and request handling. Moreover, the maintenance of the two example graphs (DODAG and spanning tree) is illustrated later in Protocol \ref{alg:p1} and Protocol \ref{alg:p2}, which are replaceable by any distributed routing algorithms.

\begin{algorithm}
\caption{Asynchronous routing scheme}
\label{alg:pf}

\begin{algorithmic}[1]
	\Statex ($N_{prev}$, $N_{next}$: the previous and next node in the path.)
	\Statex ($N_{dest}$: destination.)
	\Statex 
 
	\LeftComment{\textbf{Resource}: System Interface and Variable}
	\State $CO \gets \text{A set of classical operations}$
	\State $QO \gets \text{A set of quantum operations}$
	\State $N_c \gets \text{The current node that is running a protocol instance}$
	\Statex

	\LeftComment{\textbf{On-demand Service:} Topology Maintenance and Listener}
	\State \Call{GraphUpdate}{$QO$, $CO$, $N_c$} \Comment{Refer to Protocol \ref{alg:p1} or \ref{alg:p2}}
	\State \Call{$CO$.Listen}{`connection\_request', $N_{prev}$, $N_{dest}$}
    	\Statex

	\LeftComment{\textbf{Request Handler:} Path Navigation and Swapping}
	\Function{Navigate}{$N_{prev}$, $N_{dest}$}
		\State $N_{next} \gets \Call{NextHopDetermination}{N_c, N_{dest}}$
		\If{$N_{next}$ is not null} 
			\State bool $i \gets \Call{\textit{QO}.EntanglementSwapping}{N_{prev}, N_{next}}$
			\If{$i$ is true} 
				\State \Call{$CO$.EmitConnRequestTo}{$N_{next}$, ($N_c$, $N_{dest}$)}
			\Else \Comment{When the quantum operation fails}
    				\State \Call{GraphUpdate}{$QO$, $CO$, $N_c$}
				\State \Call{Navigate}{$N_{prev}$, $N_{dest}$}
			\EndIf
		\Else \Comment{When the current node is not in the graph}
    			\State \Call{GraphUpdate}{$QO$, $CO$, $N_c$}
			\State \Call{Navigate}{$N_{prev}$, $N_{dest}$}
		\EndIf
	\EndFunction
	\State $\Call{\textit{CO}.ConnectionHandler}{} \gets \Call{Navigate}{N_{prev}, N_{dest}}$
\end{algorithmic}

\end{algorithm}

Protocol \ref{alg:pf} is executed on each network node. To join the network, any node has to only abide by the protocol without requiring synchronization of operations and communications across all nodes. Lines 1 and 2 of Protocol \ref{alg:pf} define the interfaces for executing classical operations (i.e., $CO$) and conducting quantum operations (i.e., $QO$). Line 3 assigns the identifier of the current node running the protocol instance to $N_c$. Starting from Line 4, an on-demand service is initiated to update the instant topology continuously, based on the rules specified in the distributed graph, such as Protocol \ref{alg:p1} or \ref{alg:p2} in the subsequent subsections. Line 5 is responsible for listening to incoming connection requests. It is worth noting that this is a self-propagating entanglement swapping process, whereby each node that receives the connection request only has to focus on connecting itself with the destination node. Once a connection request arrives, Line 21 processes it by redirecting the request to the \textsc{Navigate} function, which begins at Line 6. The \textsc{Navigate} function verifies if both the current and destination nodes are present in the distributed graph. If it finds that both nodes are present in the graph, it proceeds to conduct entanglement swapping between the qubit connecting to the previous node $N_{prev}$ and the qubit connecting to the next node $N_{next}$. Assuming a successful entanglement swapping, the \textsc{Navigate} function will send a connection request to the next node, requesting the establishment of a connection between the current and destination nodes. If the entanglement swapping is unsuccessful, or either the source or the destination node is absent from the distributed graph, the node will wait for an update to the graph before attempting to navigate again. 

In short, to initiate the creation of end-to-end entanglement between a source and a destination node, the source node issues a connection request, similar to Line 11, to begin the navigation process. Each intermediate node requests the next node along the path toward the destination to perform entanglement swapping until the destination is reached. In the subsequent two subsections, we present two illustrations of the distributed graph: DODAG and spanning tree.

\subsection{DODAG protocol}
\label{section:dodag}
\begin{figure*}
  \includegraphics[width=\textwidth]{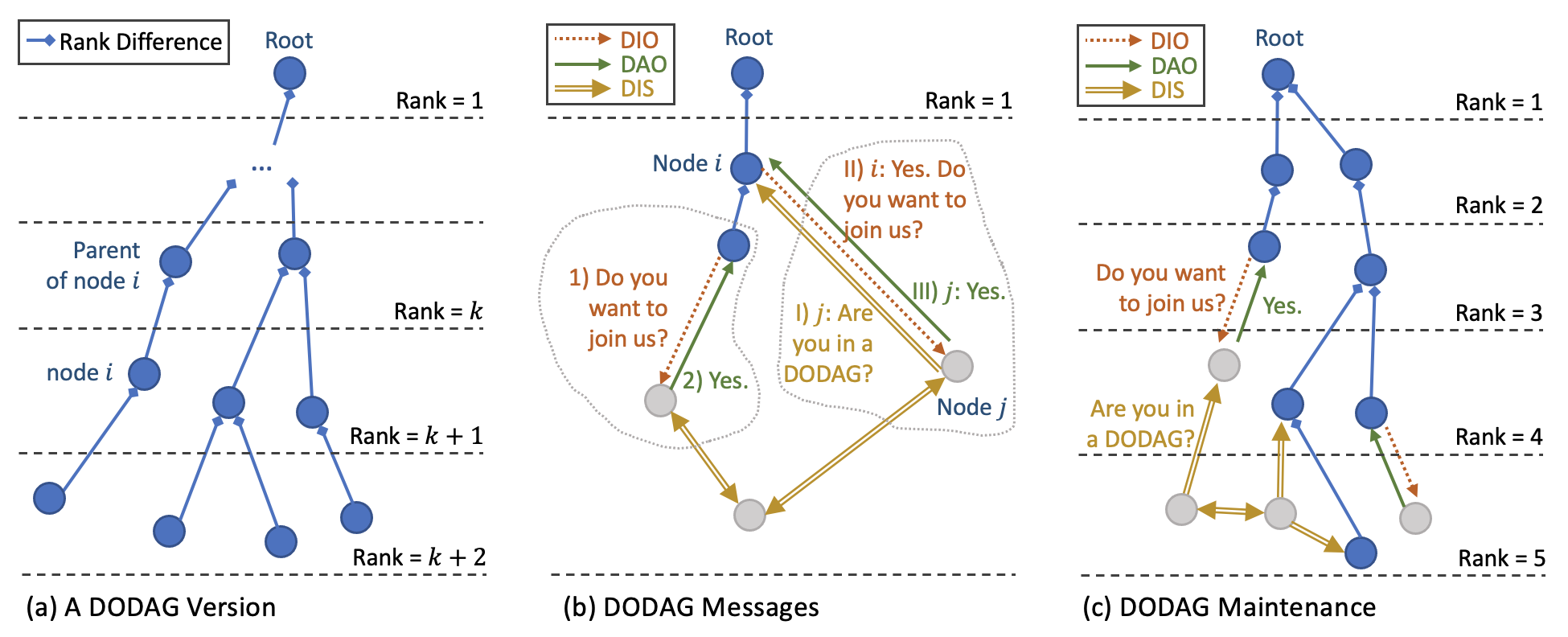}
  \captionsetup{width=1.0\linewidth}
  \caption{\textbf{DODAG and its control messages.} \textbf{(a)} This represents a version of a DODAG at any given time. The rank values of nodes in the graph increase as their distance from the root node increases. When a node is directly connected to another node and has a lower rank value, it is considered the parent of that node. Note that the rank increment here is set as 1 for simplicity, but it can be any positive real number indicating the cost metric. \textbf{(b)} Each node outside the DODAG communicates with its neighboring nodes by sending DIS messages to inquire if they belong to the DODAG. When a neighboring node responds with a yes (as Node $i$ does to Node $j$ in the figure) using DIO messages, the node (Node $j$) sends DAO messages asking to join the DODAG. Then, the neighboring node (Node $i$) computes a rank value for the new node and allows it to join. If the neighboring node does not respond, the node continues to send DIS messages until it receives a yes response or is asked to join the DODAG by a DODAG neighbor using DIO (as in the case of the dotted area on the left). \textbf{(c)} A DODAG expands itself and maintains its version using control messages indicated in (b).}
  \label{fig:dodag}
\end{figure*}

A DODAG \cite{alexander_rpl_2012} is a graph with a tree-like structure formed by selecting a root node and then growing by the exchange of control messages, as illustrated in Fig. \ref{fig:dodag}. The routing protocol revolves around the utilization of one or more DODAGs, employing designated \textit{routing metrics} and an \textit{objective function} to assist in selecting the optimal path to the root node. Subsequently, communication between the two endpoints is established via the root node. The specific graph of a DODAG at any given time is referred to as a DODAG version. If needed, a network can support multiple DODAGs with different roots.

The \textit{routing metrics} are the set of factors used to evaluate a candidate hop that determines a preferred path. These factors can vary, and here we include elements such as link quality, which is determined by link capacity ($C$) and the probability of direct-link entanglement generation ($p$), as well as node quality, which is influenced by distance to the root node, number of qubits, probability of entanglement swapping ($q$), and coherence time ($T_{co}$) — these metrics aid in creating a routing strategy that maximizes the entanglement rate. To focus on the primary concern of pathfinding, we assume that node and link qualities are homogeneous. However, in actual control messages, such as DIS (DODAG Information Solicitation), DIO (DODAG Information Object), and DAO (Destination Advertisement Object) in Fig. \ref{fig:dodag}, distinct node and link qualities may be included to assist in selecting or eliminating parents and children. For example, if a link or node falls short of the minimum requirement (e.g., fidelity $f<0.5$), it will be removed from the DODAG, along with its branches, and newer nodes will select stronger parents in their selection process. It is important to note that if a direct-link entanglement is lost (either due to decoherence or removal), the branch that was disconnected from the DODAG tree will remain a floating DODAG branch. During this time, the disconnected node will act as the temporary root node for the branch until it can successfully rejoin the main DODAG. Once it has rejoined, the floating branch will become a part of the main DODAG once again.

The \textit{objective function} for a DODAG establishes how nodes select and optimize the structure of the graph. It translates routing metrics into \textit{ranks} that determine rules such as parent selection and loop avoidance (see Appendix \ref{appe:loopavoid}). In other words, the node's rank in a DODAG indicates its connectivity with the root and is determined as a scalar measure based on the routing metrics. As the node and link qualities are considered homogenous in this paper, the ranks should increase monotonically as the nodes get farther from the root, as shown in Fig. \ref{fig:dodag}. It is worth noting that a DODAG node can have multiple parents as long as there are no loops in the graph, which is ensured by the objective function when calculating rank for new nodes. Consequently, routing can select preferred paths based on factors such as fewer hops or stronger nodes and links when dealing with distinct node and link qualities. An initial objective function is provided below for rank computation and parent selection.

Suppose a node $v_{new}$ wants to join a DODAG. It must calculate the rank step between itself and each of its neighboring nodes in the DODAG, representing the amount by which the rank should be increased along a specific link. Let us represent a neighboring DODAG node as $v_{nbr}$, and the rank step between $v_{new}$ and $v_{nbr}$ as $S(v_{new}, v_{nbr})$. The value of $S(v_{new}, v_{nbr})$ is determined by the difference between the connectivity of node $v_{new}$ (relative to the root), denoted as $\gamma(v_{new})$, and the connectivity of node $v_{nbr}$, denoted as $\gamma(v_{nbr})$. If we denote the length (i.e., the number of links) of the path from a node $v$ to the root as $L(v)$, then $\gamma(v)$ can be calculated by:
\begin{equation}
\gamma(v) = p^{L(v)}q^{L(v)-1}T_{co}
\end{equation}
With that, we can compute $\gamma(v_{nbr})$ by $L(v_{nbr})$ and compute $\gamma(v_{new})$ based on the fact $L(v_{new})=L(v_{nbr})+1$. With that, the rank step $S(v_{new}, v_{nbr})$ can be calculated as follows:
\begin{equation}
S(v_{new}, v_{nbr}) = |\gamma(v_{nbr}) - \gamma(v_{new})|
\end{equation}
The rank step can also be represented as $S(v_{new}, v_{nbr}) = |\gamma(v_{new})(\frac{1}{pq}-1)T_{co}|$ since we assume uniform $p$, $q$ and $T_{co}$. A stronger node (i.e., a node with higher connectivity) would have a smaller rank step. Next, node $v_{new}$ selects the neighboring node with the minimum rank step (suppose it is $v_{nbr}$) to become its parent node. This enables node $v_{new}$ to determine its rank value by following $v_{nbr}$, denoted as $\rho(v_{new})$, when it successfully joins the DODAG:
\begin{equation}
\rho(v_{new}) = \rho(v_{nbr}) + S(v_{new}, v_{nbr})
\end{equation}

While a DODAG is constantly updated according to the rules above (refer to Appendix \ref{appe:loopavoid} for information on loop avoidance), the routing process in the DODAG is as simple as moving up to the root and then descending to the intended destination by selecting the parent with the lowest rank value among all parents (or both source and destination ascend to the root). The maintenance of the DODAG is outlined in Protocol \ref{alg:p1}.

\renewcommand{\thealgorithm}{1.\arabic{algorithm}}
\setcounter{algorithm}{0}
\begin{algorithm}[b]
\caption{DODAG maintenance}\label{alg:p1}
\begin{algorithmic}[1]
	\Statex ($N_{neighbor}$: the node that the current node is communicating with)
	\Function{NextHopDetermination}{$N_c$, $N_{dest}$}
		\State \Return $N_c$.parent
	\EndFunction
	\Function{GraphUpdate}{$QO$, $CO$, $N_c$}
		\State $\Call{\textit{CO}.DIS\_Handler}{} \gets \Call{DIS}{N_{neighbor}, CO}$
		\State $\Call{\textit{CO}.DIO\_Handler}{} \gets \Call{DIO}{N_{neighbor}, QO, CO}$
		\While {true}
			\If{$N_c$ is in DODAG}
				\State \Call{$CO$.BroadcastDIO}{$N_c$}
				\State \Call{$CO$.Listen}{`DIS', $N_{neighbor}$}
			\Else
				\State \Call{$CO$.BroadcastDIS}{$N_c$}
				\State \Call{$CO$.Listen}{`DIO', $N_{neighbor}$}
			\EndIf
		\EndWhile
	\EndFunction
	\Function{DIS}{$N_{neighbor}$, $CO$}
		\State \Call{$CO$.SendDIO}{$N_{neighbor}$}
	\EndFunction
	\Function{DIO}{$N_{neighbor}$, $QO$, $CO$}
		\State bool $j \gets \Call{\textit{CO}.CheckMiniRank}{}$
		\If{$j$ is true}
			\State \Call{$CO$.SendDAO}{$N_{neighbor}$}
			\State bool $k \gets \text{false}$
			\While {$!k$}
				\State $k \gets \Call{\textit{QO}.EntanglementGeneration}{N_{c}, N_{neighbor}}$
			\EndWhile
		\EndIf
	\EndFunction
\end{algorithmic}
\end{algorithm}

\subsection{Distributed spanning tree}
\label{section:dst}
As a result of having access solely to local knowledge of the instant topology of a quantum network, constructing a spanning tree — a subgraph of a connected undirected graph that covers all vertices using the least number of edges — necessitates a distributed approach. Numerous distributed algorithms for spanning tree construction have been developed, including the renowned Gallagher-Humblet-Spira (GHS) algorithm \cite{gallager_distributed_1983}. At the beginning of a GHS algorithm, each node is considered a fragment, and these fragments progressively fuse to form larger ones. To combine with another fragment, a fragment must locate its least-weight outgoing edge and work collaboratively with other nodes to find it in a distributed fashion. Over time, the number of fragments reduces until only one fragment remains, which is the minimum spanning tree (MST). A spanning tree guarantees the existence of a path between any two nodes. Nevertheless, using a path will disconnect parts of the tree, requiring all nodes to update it continuously. Our assumption of uniform $p$, $q$, and $T_{co}$ implies that all edge weights are equal in our situation. 

\begin{figure*}
  \includegraphics[width=\textwidth]{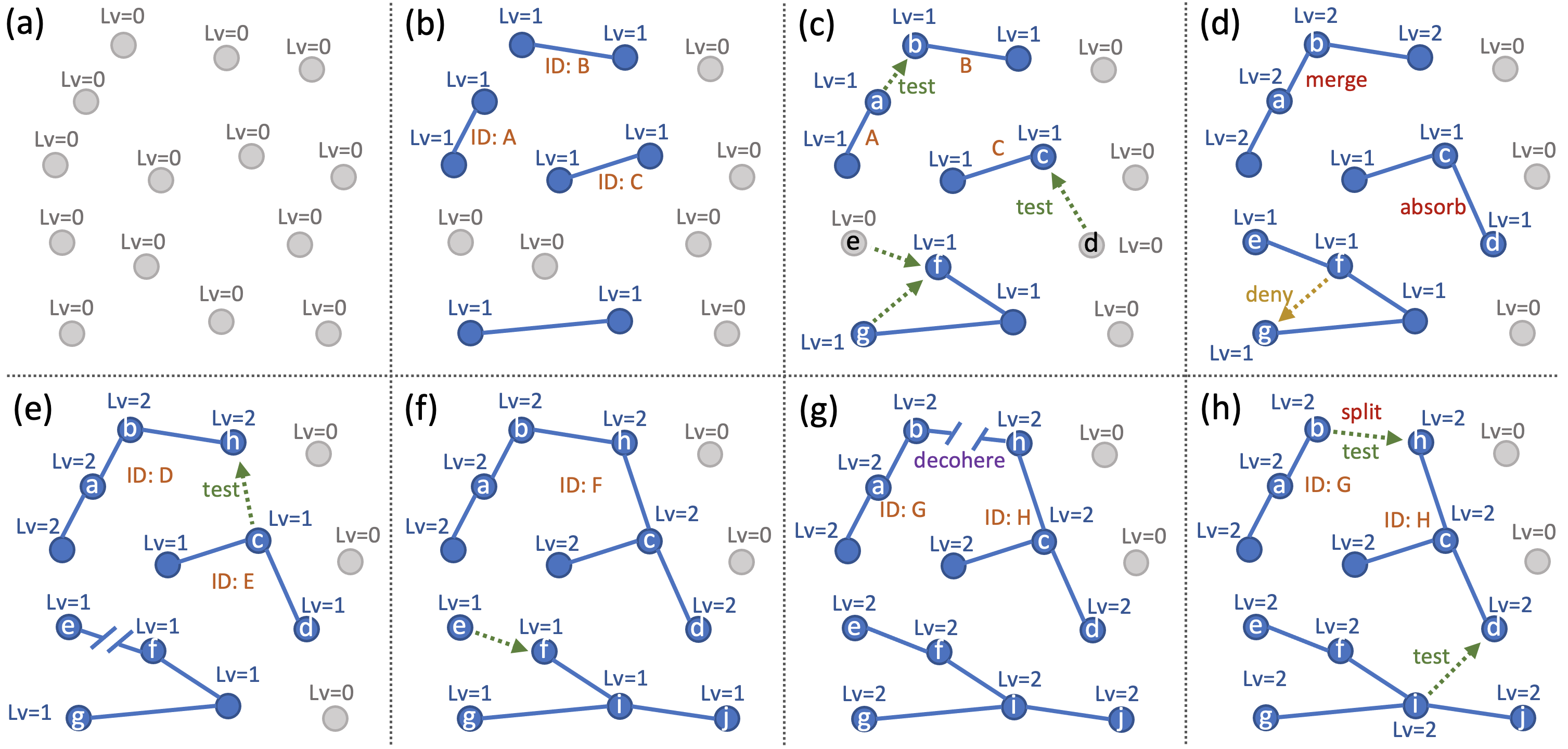}
  \captionsetup{width=1.0\linewidth}
  \caption{\textbf{Distributed spanning tree by the modified GHS algorithm.} The figure illustrates the maintenance of instant topology in a stepwise manner as a spanning tree without including the physical links. \textbf{(a)} Initially, all nodes are considered as fragments with a level value of $0$. \textbf{(b)} Each node sends a "test" message to its adjacent nodes, resulting in some successful and some failed connections (i.e., entanglement links). Since all level values are the same at the start, any successful connection leads to a merge process where the levels are incremented by $1$, and the connecting edge becomes the core edge of that fragment. The fragments are denoted as $A$, $B$, and $C$. \textbf{(c)} Nodes connected by core edges send "test" messages to their adjacent nodes (not every "test" message is shown for brevity). For example, Node $a$ sends a message to Node $b$, and Node $d$ sends a message to Node $c$. \textbf{(d)} Fragments $A$ and $B$ have the same level, so they merge, and their levels are incremented by $1$. Fragment $C$ absorbs another node, and its level remains the same. Note that Node $f$ declines to connect with Node $g$ because they have the same Fragment ID. \textbf{(e)} Node $c$ sends a "test" message to Node $h$. \textbf{(f)} Fragment $D$ absorbs Fragment $E$ and becomes Fragment $F$. \textbf{(g)} Edge $b-h$ breaks, causing Fragment $F$ to split into Fragments $G$ and $H$. \textbf{(h)} The split event is triggered, and Nodes $b$ and $h$ send "test" messages to their adjacent nodes.}
  \label{fig:spanningtree}
\end{figure*}

However, we must address the instant topology's probabilistic and decoherent entanglements (links). Fig. \ref{fig:spanningtree} illustrates our modified version of the GHS algorithm. In short, it addresses the separation of a fragment caused by the decoherence of entanglements and continuous attempts to generate an outgoing edge due to the probabilistic nature of entanglement generation. We achieve this by incorporating a \textit{split} event in addition to the \textit{merge} and \textit{absorb} processes of the GHS algorithm. The following paragraphs outline how the modified GHS algorithm operates within a quantum network.

In the physical topology $G(V, E)$, each node starts as a \textit{fragment}, and the objective is to connect all nodes with minimum weight (i.e., the least number of edges in our case) with no loop. In this context, an edge represents a direct-link entanglement in a quantum network. A fragment $F$ is a subset of $V$ and is assigned a non-decreasing integer level $Lv(F)$ that begins with $0$. Each node in a fragment must have the same level value as its fragment, i.e., $Lv(v) = Lv(F)$ $\forall$ $v \in F$. Fragments with a non-zero level are assigned a unique name or fragment ID, denoted as $FID(v) = FID(F)$ $\forall$ $v \in F$. Every node chooses a neighbor node to form a core edge under certain constraints, such as the neighbor node with the minimum weight. Each fragment has only one core edge, and the core edge selection process is beyond this text's scope but can be found in Reference \cite{gallager_distributed_1983}. The two nodes connected by a core edge are responsible for broadcasting the fragment ID and level to all other nodes in the fragment. After receiving a broadcast message from the core, every node searches for its minimum-weight outgoing incident edge. For instance, if node $v$ receives a broadcast, it selects its minimum weight edge and sends a "test" message to node $v'$ on the other side (assuming $v' \in F'$) with its fragment's ID and level. Then, node $v'$ determines whether the edge is an outgoing edge and sends a message back to inform node $v$ of the result. The decision is based on the following criteria: 
\begin{enumerate}
  \item if $FID(v) = FID(v')$, the edge is not outgoing because they are in the same fragment;
  \item if $FID(v) \neq FID(v')$ and $Lv(v) \leq Lv(v')$, the edge is outgoing;
  \item if $FID(v) \neq FID(v')$ and $Lv(v) > Lv(v')$, the algorithm cannot reach a conclusion because the two nodes may already belong to the same fragment, but node $v'$ has not yet discovered this due to the delay of a broadcast message. In this case, the algorithm allows node $v'$ to postpone its response until its level becomes higher than or equal to the level it received from node $v$. Refer to Appendix \ref{appe:ghsdelay} for scenarios involving broadcast delay in the GHS algorithm.
\end{enumerate}

When an outgoing edge is present, the fragments will begin to merge through the \textit{merge} and \textit{absorb} processes (direct-link entanglement generation on the outgoing edge). If $Lv(F) = Lv(F')$, the fragments will merge through the merge process, resulting in a new fragment with an increased level of $Lv(F)+1$. On the other hand, if $Lv(F) < Lv(F')$, the fragments will merge through the absorb process, and the combined fragment will have the same level as $F'$. When an edge within a fragment decoheres, it triggers a \textit{split} event, where the two adjacent nodes generate new fragment IDs for their own fragments and broadcast them to the rest of the fragment. The levels of the new fragments remain the same. The increase in level during the merge process is to prevent the formation of an edge to a fragment that has already been combined by the other outgoing edge, which can cause a loop (see Appendix \ref{appe:ghsdelay}). The split event does not combine fragments and thus does not affect the merge and absorb processes or the levels of any fragments. 

As the rules dictate, a distributed spanning tree undergoes continuous updates. During the routing process within the spanning tree, the objective is to determine the correct direction toward the destination from the multiple available directions. Whenever an outgoing edge is detected, the relative position of each node in the tree, along with the new level and fragment ID, will be broadcast. This relative position is used to determine the correct direction toward the destination. Protocol \ref{alg:p2} provides an outline of the distributed spanning tree's update process. To be concise, we have omitted the broadcasting and listening of events that update the fragment ID and level within a fragment.

\begin{algorithm}
\caption{Distributed spanning tree}\label{alg:p2}
\begin{algorithmic}[1]
	\Statex ($N_{neighbor}$: the node that the current node is communicating with)
	\Function{NextHopDetermination}{$N_c$, $N_{dest}$}
		\State \Return $N_c$.next \Comment{relative position is updated when receiving a broadcast}
	\EndFunction
	\Function{GraphUpdate}{$QO$, $CO$, $N_c$}
		\State $\Call{\textit{CO}.TestMessage\_Handler}{} \gets \Call{TestMessage}{N_{neighbor}, QO, CO}$
		\While {true}
			\State \Call{$CO$.BroadcastTestMessage}{$N_c$}
			\State \Call{$CO$.Listen}{`test', $N_{neighbor}$}
		\EndWhile
	\EndFunction
	\Function{TestMessage}{$N_{neighbor}$, $QO$, $CO$}
		\If{$CO.FID(N_{neighbor}) != CO.FID(N_{neighbor})$}
			\If{$CO.Lv(N_{c}) \leq CO.Lv(N_{neighbor})$}
				\State bool $r \gets \text{false}$
				\While {$!r$}
					\State $r \gets \Call{\textit{QO}.EntanglementGeneration}{N_{c}, N_{neighbor}}$
				\EndWhile
				\If{$CO.Lv(N_{c}) == CO.Lv(N_{neighbor})$} \Comment{merge}
					\State char $id \gets \Call{CO.GenerateID}{}$
					\State int $lv \gets CO.Lv(N_{c})+1$
					\State \Call{$CO$.SetID}{$N_{c}$, $id$}
					\State \Call{$CO$.SetID}{$N_{neighbor}$, $id$}
					\State \Call{$CO$.SetLv}{$N_{c}$, $lv$}
					\State \Call{$CO$.SetLv}{$N_{neighbor}$, $lv$}
				\Else \Comment{absorb}
					\State char $id \gets CO.FID(N_{neighbor})$
					\State int $lv \gets CO.Lv(N_{neighbor})$
					\State \Call{$CO$.SetID}{$N_{c}$, $id$}
					\State \Call{$CO$.SetLv}{$N_{c}$, $lv$}
				\EndIf
				\State \Call{$CO$.BroadcastNewFragementInfo}{$N_{c}$, $id$, $lv$}
				\State \Call{$CO$.BroadcastNewFragementInfo}{$N_{neighbor}$, $id$, $lv$}
			\Else
				\State \Call{TestMessage}{$N_{neighbor}$, $QO$, $CO$}
			\EndIf
		\EndIf
	\EndFunction
\end{algorithmic}
\end{algorithm}

\subsection{Time complexity}
\label{section:timecomp}
The time complexity of asynchronous pathfinding algorithms can vary depending on the specifics of the algorithm and the nature of the network. Here, we analyze the time complexity of the asynchronous protocols in terms of link availability probability $p$, node availability probability $q$, and consider a grid network with node size $n$ (i.e., $\sqrt{n} \times \sqrt{n}$ grid) where the root is at the center of the grid.

The time complexity of DODAG can be analyzed in two parts: the time complexity for constructing the DODAG and the time complexity of pathfinding within the DODAG. In the worst case, constructing a DODAG from a network can have a time complexity of $O(n)$ if we consider all possible nodes and links. However, since each link and node has a probability $p$ and $q$ of being available, respectively, on average, we are dealing with $p\sqrt{n}$ nodes and $q\sqrt{n}$ links (assuming a full mesh). So, the complexity becomes $O(pqn)$ for creating the DODAG. However, this decreases the probability of the pathfinding part of the protocol. In other words, the reduction of time complexity of DODAG construction of the algorithms (by the decrease of $p$ and $q$) increases the success probability of pathfinding. For pathfinding, the paths from the source or destination to the root are planned by the rank values, which means the complexity corresponds to the length of the planned path. The worst-case time complexity of DODAG pathfinding is thus the summation of the furthest possible distance between source (i.e., $\sqrt{n}$) and root and that between destination and root (i.e., $\sqrt{n}$). Therefore, the overall complexity is $O(\sqrt{n})$. Note that these are the worst-case time complexities of the algorithm since we consider DODAG construction and pathfinding as separate steps, but they are happening simultaneously in practice. Thus, the true complexity can be dynamic and less than this in practice.

The time complexity of the spanning tree protocol can be similarly analyzed in two parts: constructing the distributed spanning tree and pathfinding in the spanning tree. Generally, a distributed algorithm to construct a minimum spanning tree (like the GHS algorithm) can have a time complexity of $O(n \log n)$ \cite{gallager_distributed_1983}. In the worst case, all fragments are combined with "Merge" operations, so the number of fragments decreases by half at each level. Therefore, the maximum number of levels is $O(\log n)$. Each level takes $O(n)$ time (for information exchange in broadcast). Hence, the time complexity is $O(n \log n)$. Since a spanning tree is a tree, and trees are acyclic and have only one path between any two nodes, locating a path in a tree (given pointers to the start and end nodes) can be done in $O(n)$ in the worst case, as you may need to traverse the entire tree (in a skewed tree scenario). But if it is reasonably balanced, the path could be found much faster, in $O(\log n)$ time. Note that the discussion of message complexity of both algorithms is out of the scope of this article but can be found in \cite{korach_tight_1984}.

\section{Performance analysis}
\begin{figure*}
  \includegraphics[width=\textwidth]{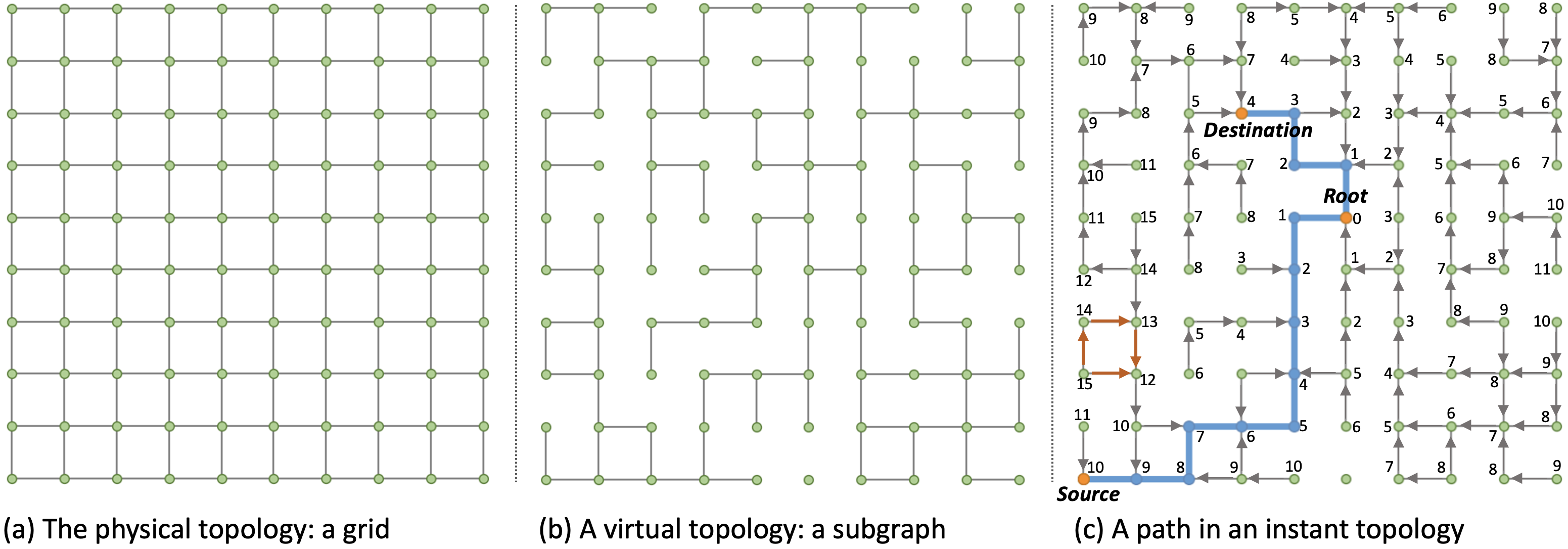}
  \captionsetup{width=1.0\linewidth}
  \caption{\textbf{Grid topology.} \textbf{(a)} A 2D grid represents the physical topology $G(V, E)$. \textbf{(b)} An instant topology $G'(V', E')$ is a subset of this grid. \textbf{(c)} A DODAG path can be identified within an instant topology, with the connected graph forming a DODAG with respect to the root node. The arrows indicate the rank differences (parent-child relationships), while small digits beside nodes indicate their rank value. It is important to note that the small squares in (b) and (c) do not represent loops but occur when nodes select multiple parents, as shown in the example of orange arrows in (c). These integers are solely to provide an example that illustrates the relative difference of rank values. Real values would be more fine-grained (i.e., can be any positive real numbers).}
  \label{fig:grid}
\end{figure*}

To assess the effectiveness of our asynchronous routing protocol, we utilize the entanglement rate ($\xi$) as a metric. To simplify matters, we focus on a 2D grid topology (as illustrated in Fig. \ref{fig:grid}a), denoted by $G(V, E)$, and define an instant topology $G'(V', E')$ as a subset of this grid. Each node in $G$ is only aware of the status of entanglements in the four direct links connected to it and has no knowledge of the larger instant topology beyond that.

In Section \ref{section:dodag}, we used $L(v)$ to denote the length (number of links) of a path from node $v$ to the root of a DODAG. It is possible to extend this notation to represent the length of the path between any two nodes (e.g., source node $s$ and target node $t$) as $L(s,t)$, but this value only applies to a specific instance of the instant topology. To calculate the entanglement rate in a more general case, we can estimate this value by approximating the mean length of all paths between two nodes in the physical topology. Specifically, we use $l_{s,t}$ to represent the mean length of all possible paths between nodes $s$ and $t$ in $G$. $l_{s,t}$ is of interest in percolation theory for a 2D square lattice, but it cannot be determined analytically. However, an approximation is possible \cite{malik_concurrence_2022, broadbent_percolation_1957} (see Appendix \ref{appe:meanpath}).

\subsection{Synchronous protocol analysis}
Let us denote the average number of disjoint paths found between $s$ and $t$ as $n_{s, t}$, and in a 2D grid topology, we have $0 \leq n_{s, t} \leq 4$ since the maximum number of neighboring links that can be used by a node is four. Assuming we have only one possible entanglement in a direct link (i.e., $C/E(f)=1$), we can estimate the average entanglement rate $\xi_{syn}$ (per time slot) between nodes $s$ and $t$ for existing synchronous protocols as follows:
\begin{equation}
\label{eqn:xis}
\xi_{syn}(s, t) = p^{l_{s,t}} q^{l_{s,t}-1} n_{s, t}
\end{equation}
Let us denote the length of the shortest path from $s$ to $t$ as $\hat{l}_{s, t}$. With the range of $n_{s, t}$, we can establish the bounds for $\xi_{syn}(s, t)$ (the length of the shortest path is the minimum length for each of the four paths):
\begin{equation}
0 \leq \xi_{syn}(s, t) \leq 4p^{\hat{l}_{s,t}} q^{\hat{l}_{s,t}-1}
\end{equation}
It is noteworthy that our examination of the entanglement rate pertains to exclusive connection requests. For an analysis of link capacity allocation in the context of a batch of requests, we refer interested readers to Reference \cite{li_effective_2021}.

\subsection{Asynchronous protocol analysis}
\label{section:dodagana}
In essence, the idea of asynchronous protocols is to preserve the direct-link entanglements that do not fall on the route between the two communicating parties, thus enabling the upkeep of a distributed graph such as DODAG and a spanning tree for future use. Specifically, we take the DODAG protocol as an example and examine its performance in this subsection. Besides $s$ and $t$, the root node of the DODAG is denoted by $r$. We define $l'_{s, r}$ as the average number of direct-link entanglements already in the path between the root node $r$ and node $s$, and $l'_{r, t}$ as that between nodes $r$ and $t$. Note that $l'_{s, r}$ and $l'_{r, t}$ are proportional to $T_{co}$ because higher values of $T_{co}$ result in more entanglement links that remain un-decohered in $G'$. With this, we can estimate the entanglement rate $\xi_{DODAG}$ for the DODAG protocol as:
\begin{equation}
\label{eqn:xidodag}
\xi_{DODAG}(s, t) = p^{l_{s,r}+l_{r,t}-l'_{s, r}-l'_{r, t}} q^{l_{s,r}+l_{r,t}-1} n_{s, t} 
\end{equation}
To facilitate comparison, we adopt a unit time for the entanglement rate that is equivalent to one time slot in synchronous approaches. As mentioned earlier, we define the unit time as $T = \frac{T_{co}}{n}$, where $n \in \mathbb{Z}$ and $n \ge 1$. For brevity, we disregard the symbol $T$ and express $T_{co}$ as $n$ unit times. Since $l'_{s, r} \propto T_{co}$ and $l'_{r, t} \propto T_{co}$, it follows that $\xi_{DODAG}(s, t)$ increases as the coherence time grows. Nonetheless, by definition, we know that $\xi_{syn}(s, t)$ is independent of $T_{co}$.

The range of $n_{s,t}$ in our asynchronous approach is the same as in the synchronous approach, as a node can have up to four paths to the destination (having four parent nodes). Let us denote the length of the shortest path from $s$ to $r$ and from $r$ to $t$ as $\hat{l}_{s, r}$ and $\hat{l}_{r, t}$, respectively. Then, we can establish the bounds of $\xi_{DODAG}(s,t)$ as follows (assuming $s$ and $t$ need to go through the root $r$):
\begin{equation}
\label{eqn:ub}
0 \leq \xi_{DODAG}(s, t) \leq 4q^{\hat{l}_{s, r}+\hat{l}_{r, t}-1}
\end{equation}
The upper bound is attained when both $s$ and $t$ are already present in the DODAG, meaning no direct-link entanglement generation (expressed by $p$) is necessary. It should be noted that our assumption for Equation \ref{eqn:ub} is that $s$ and $t$ must pass through the root node $r$. However, in reality, if $s$ and $t$ are located on the same branch, they can pass through a shared ancestor instead. As such, the right-hand side of Equation \ref{eqn:ub} represents a worst-case upper bound, and in reality, the upper bounds for most specific scenarios would be greater. We demonstrate that even in the worst-case scenario where nodes need to traverse the root node, the DODAG protocol provides a higher upper bound (for most node pairs $s$ and $t$) than existing synchronous approaches. On the other hand, the lower bound happens when there are no direct-link entanglements in $G'$, and $s$ and $t$ cannot be incorporated into the DODAG within a single unit time.

\begin{figure}
  \includegraphics[width=0.7\textwidth]{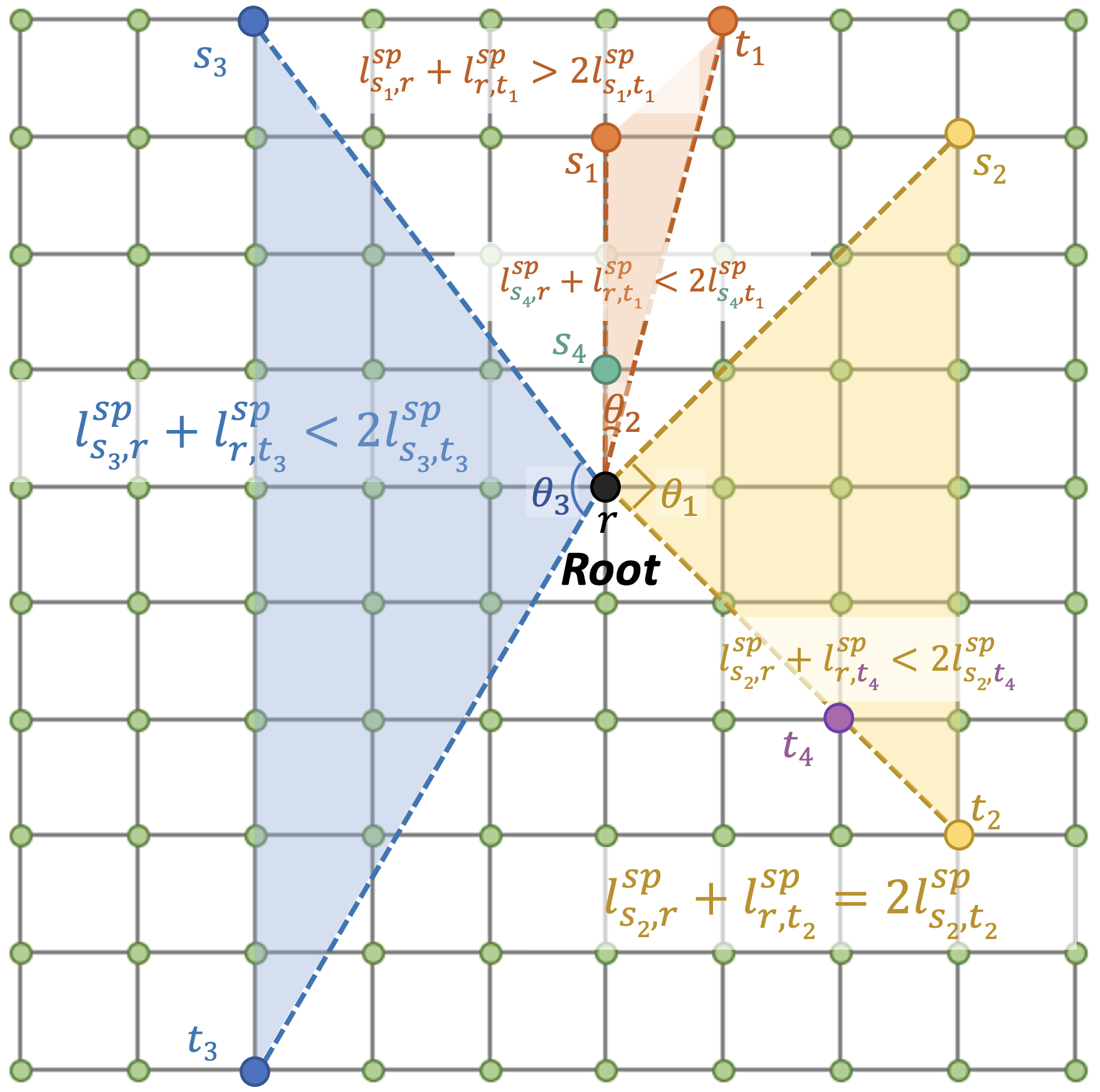}
  \caption{\textbf{The location scenarios in a DODAG.}}
  \label{fig:dodagana}
\end{figure}

The worst-case upper bound of our approaches is represented by $\xi^{UB}_{DODAG}(s, t) = 4q^{\hat{l}_{s, r}+\hat{l}_{r, t}-1}$. Initially, let us examine the case where $p=q$, then the upper bound of synchronous approaches is $\xi^{UB}_{syn}(s, t) = 4q^{2\hat{l}_{s,t}-1}$. Therefore, the difference between $\xi^{UB}_{DODAG}(s, t)$ and $\xi^{UB}_{syn}(s, t)$ is attributed to the difference between $2\hat{l}_{s,t}$ and $\hat{l}_{s, r}+\hat{l}_{r, t}$. Let us denote the angle formed by $s$, $r$, and $t$ with respect to $r$ as $\theta$, for instance, $\theta_1$, $\theta_2$, and $\theta_3$ in Fig. \ref{fig:dodagana}. As demonstrated in Fig. \ref{fig:dodagana}, we can infer the difference between $2\hat{l}_{s,t}$ and $\hat{l}_{s, r}+\hat{l}_{r, t}$ by the relationship of sides in a triangle:
\begin{enumerate}
  \item If $\theta$ is obtuse (e.g., $\theta_3$), $\hat{l}_{s, r}+\hat{l}_{r, t} < 2\hat{l}_{s,t}$ (i.e., the sum of the two sides adjacent to an obtuse triangle is less than twice the length of the other side. See the next subsection).
  \item If $\theta$ is a right angle and the three nodes form an isosceles right triangle (e.g., $\theta_1$ with $s_2$ and $t_2$), $\hat{l}_{s, r}+\hat{l}_{r, t} = 2\hat{l}_{s,t}$. If $\theta$ is a right angle and the three nodes do not form an isosceles right triangle (e.g., $\theta_1$ with $s_2$ and $t_4$), $\hat{l}_{s, r}+\hat{l}_{r, t} < 2\hat{l}_{s,t}$. 
  \item If $\theta$ is acute (e.g., $\theta_2$), there are two cases depending on the relative position of the three nodes. Assuming $s$ is the one closer to $r$ (i.e., $t$ is farther from the root. Note that these are also valid when treating $t$ as the closer one to the root):
  \begin{enumerate}
    \item When $\hat{l}_{s, r} \leq \alpha \hat{l}_{s, t}$ (See the next subsection for the value of $\alpha$ with respect to $\theta$), $\hat{l}_{s, r}+\hat{l}_{r, t} \leq 2\hat{l}_{s,t}$ (e.g., $\theta_2$ with $s_1$ and $t_1$). That is, our approach is better when $s$ is close to the root but far from $t$. In the case of equality, the three nodes form an equilateral triangle.
    \item When $\hat{l}_{s, r}>\alpha \hat{l}_{s, t}$, $\hat{l}_{s, r}+\hat{l}_{r, t} < 2\hat{l}_{s,t}$ (e.g., $\theta_2$ with $s_4$ and $t_1$). That is, when $s$ is close to $t$ but far from the root, our approach is not providing the optimal route.
  \end{enumerate}
\end{enumerate}

In summary, only if the source and destination nodes are located in proximity and far from the root (only case 3.b above), our worst-case upper bound is lower than $\xi^{UB}_{syn}(s, t)$. Nevertheless, in this scenario, they are likely to be on the same branch of the DODAG and thus have a higher probability of having a much greater rate since they do not need to pass through the root. This hypothesis is verified through simulation in the subsequent section. Additionally, in the case where the three nodes are aligned in a line, if $r$ is between $s$ and $t$, $\hat{l}_{s, r}+\hat{l}_{r, t} = \hat{l}_{s,t}$. If $r$ is not in between, they can always establish communication without going through the root. When traveling towards the root, the node farther away from the root would reach the closer node before reaching the root. However, any entanglement swapping between the closer node and the root would be pointless. Therefore, we can apply $\hat{l}_{s, r}+\hat{l}_{r, t} = \hat{l}_{s,t}$ in this scenario as well.

With that, it is obvious that $\xi^{UB}_{DODAG}(s, t) > \xi^{UB}_{syn}(s, t)$ holds when $p<q$. 

Now, let us see the case where $p>q$. From $\xi^{UB}_{syn}(s, t)$ and $\xi^{UB}_{DODAG}(s, t)$, we have the relationship between $\hat{l}$ and $\hat{l}_{s, r}+\hat{l}_{r, t}$:
\begin{enumerate}
    \item If $\hat{l}_{s,t} \geq \beta(\hat{l}_{s, r}+\hat{l}_{r, t})$ where $\beta<1$ (See the next subsection for the value of $\beta$ with respect to $p$ and $q$), $\xi^{UB}_{syn}(s, t) \leq \xi^{UB}_{DODAG}(s, t)$.
    \item If $\hat{l}_{s,t} < \beta(\hat{l}_{s, r}+\hat{l}_{r, t})$, $\xi^{UB}_{syn}(s, t) > \xi^{UB}_{DODAG}(s, t)$.
\end{enumerate}

\begin{figure*}
  \includegraphics[width=\textwidth]{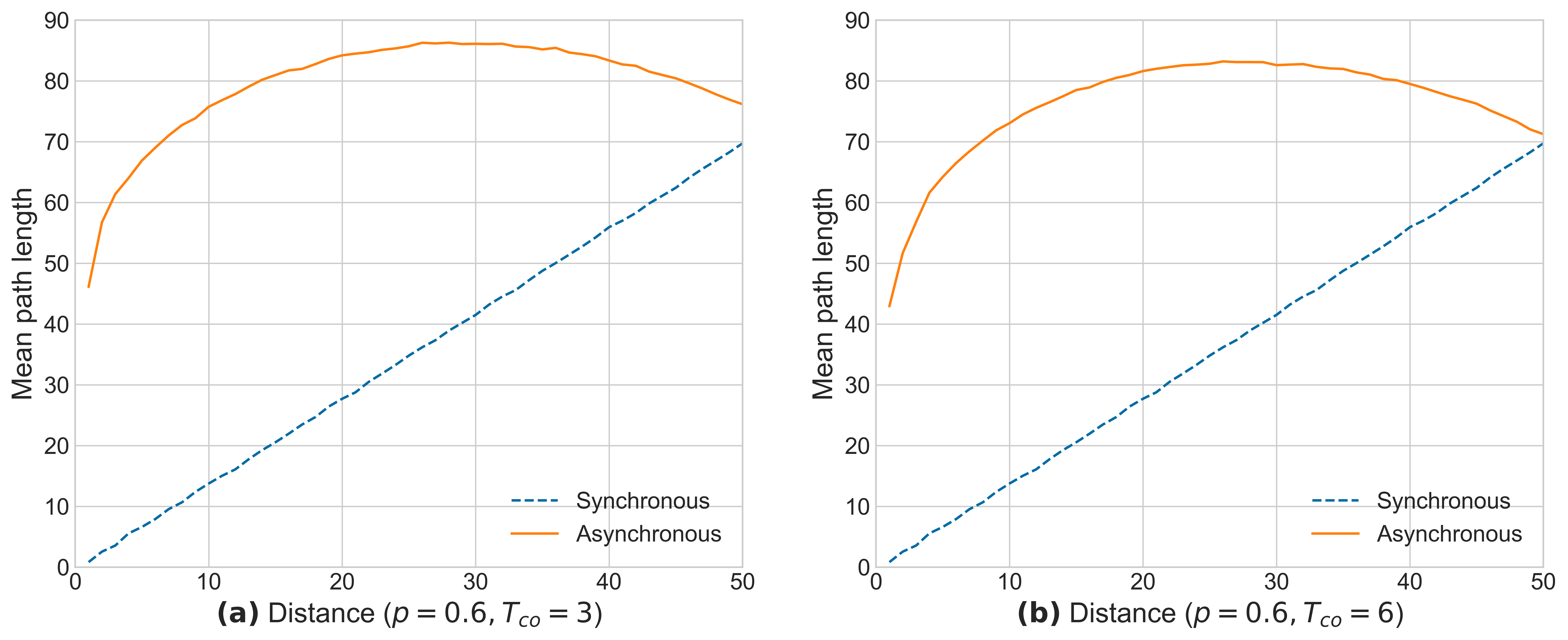}
  \captionsetup{width=1.0\linewidth}
  \caption{\textbf{Mean path length when source and destination must pass through the root.} \textbf{(a)} With $p=0.6, T_{co}=3$ (with source and destination locations randomly generated), the average number of hops in a path in an asynchronous approach increases rapidly when the L1 distance (see Appendix \ref{appe:meanpath}) between source and destination is small and becomes gentle. When the distance further increases, the mean hops decrease gradually. This happens because the two nodes and the root tend to form a line (setting the root in the center of the network) as the distance between the two communicating nodes increases. It is important to note that although the mean path length tends to decrease with larger distances between nodes, it will not be lower than the actual distance for obvious reasons. This trend scales as the number of network nodes increases, and at the maximum distance between the source and destination, the mean number of path lengths for both approaches always converges. In this figure, the network size is $26\times26$, where the maximum distance is $50$. Note that the value of $q$ is irrelevant in this context as the graph solely illustrates the pathfinding in an instant topology without yet reaching the phase of implementing entanglement swapping. \textbf{(b)} The plot with the value of $T_{co}$ set to 6.}
  \label{fig:hopsdistance}
\end{figure*}

Even though the second case above seems to suggest that the synchronous method has a higher upper bound on the entanglement rate, the above relationships are valid only because $\hat{l}_{s, r}+\hat{l}_{r, t} \geq  \hat{l}_{s,t}$. However, we find that, in reality, it is most likely $\hat{l}_{s, r}+\hat{l}_{r, t} \approx  \hat{l}_{s,t}$ when the distance between the source and destination is large. As shown in our simulation in Fig. \ref{fig:hopsdistance}, the paths' lengths (hops) tend to converge as the distance between the source and destination increases. That is, $l_{s, r}+l_{r, t} \approx l_{s,t}$ when the distance between source and destination is large, which means that our approach has a significantly larger upper bound than asynchronous approaches when the distance between source and destination is large. This phenomenon can be explained by Fig. \ref{fig:dodagana} to some extent (i.e., as the distance between the two communicating nodes increases, the two nodes and the root tend to align in a linear configuration if the root is in the center of the network, that is, the three nodes are more likely to form a line as distance increases. It is important to note that although the mean path length tends to decrease with larger distances between nodes, it will not be lower than the actual distance for obvious reasons. The number of network nodes influences the observed trend in a grid network. When approaching the maximum achievable distance between the source and destination, the mean path lengths for both approaches always converge. This convergence is demonstrated in Fig. \ref{fig:hopsdistance} (a) and (b).). Also, this distance value of the convergence point becomes smaller when $T_{co}$ becomes larger. With this fact, Equations \ref{eqn:xis} and \ref{eqn:xidodag} imply $\xi_{DODAG}(s, t) \gtrapprox \xi_{syn}(s, t)$ when the distance between source and destination is large, which makes $l_{s, r}+l_{r, t} \approx l_{s,t}$. 

Note that Fig. \ref{fig:hopsdistance} only represents the average length of paths found by the two protocols in an instant topology. It does not take into account the probability of finding a path. Path length alone does not fully determine the entanglement rate. We must also consider the probability of finding a path in an instant topology to determine the entanglement rate. In our simulations, the number of successful path findings by asynchronous protocols is significantly higher than synchronous protocols, given the same number of simulation iterations. The averages in Fig. \ref{fig:hopsdistance} are calculated by taking only the iterations with successful path findings. Therefore, despite potentially having longer paths, asynchronous protocols are more likely to find a path, leading to a higher entanglement rate in the end. Results pertaining to entanglement rate versus distances are presented in the next subsection.

Additionally, the simulations presented in Fig. \ref{fig:hopsdistance} only account for scenarios where the source and destination nodes must pass through the root to establish a connection. However, in reality (and in subsequent simulations), it is common for closely located source and destination nodes to bypass the root and instead traverse a common ancestor. As a result, the mean path length for our approach is expected to be considerably lower than depicted in the figure (as demonstrated in the simulation results in Subsection \ref{section:simu_res}).

Moreover, if we assume that there is at least one available path (where $p$ and $q$ are sufficiently high to exceed the probability threshold in bond percolation, such as $p>0.5$ and $q=1$ for a 2D square lattice \cite{kesten_critical_1980}), we can establish lower bounds for $\xi^{LB}_{syn}(s, t) = p^{l_{s,t}} q^{l_{s,t}-1}$ and $\xi^{LB}_{DODAG}(s, t) = p^{l_{s, r}+l_{r, t}} q^{l_{s, r}+l_{r, t}-1}$. If no direct-link entanglement is available in the instant topology, using a DODAG approach results in a worse lower bound since it requires two paths to achieve, the sum of which is often larger than the shortest path achievable in the existing synchronous approach. However, as the distance between the source and destination increases (network scale increases), the two approaches have similar lower bounds.

\subsection{Relative positions of the three nodes in DODAG}
\label{subsection:rel_pos}
\begin{figure}
  \includegraphics[width=0.8\textwidth]{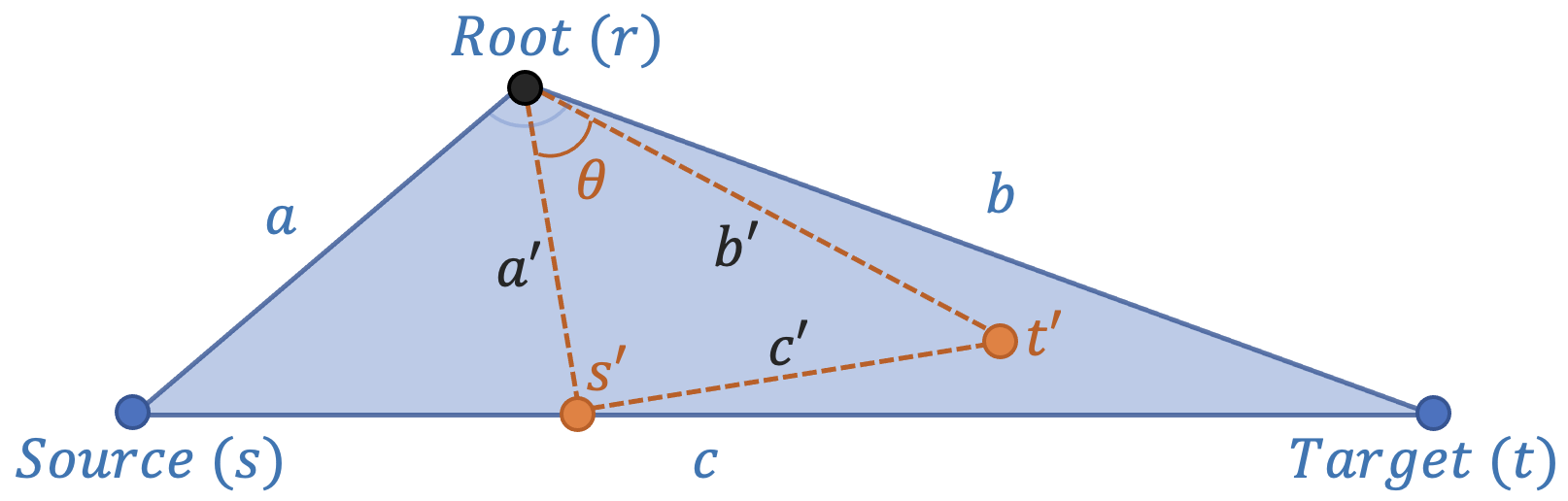}
  \captionsetup{width=1.0\linewidth}
  \caption{\textbf{The triangle formed by $r$, $s$, $t$.} To simplify the symbols, we denote the sides of the triangle formed by the three nodes as $a$, $b$, and $c$.}
  \label{fig:triangle_rst}
\end{figure}

\subsubsection{Sides of an obtuse triangle}
\label{subsubsection:sides}
As shown in Fig. \ref{fig:triangle_rst}, let us denote the sides of the triangle formed by the three nodes as $a$, $b$ and $c$. We show that the sum of the two sides adjacent to the obtuse angle of an obtuse triangle is less than twice the length of the third side. Since it is an obtuse triangle and $c$ is the longest side, we have 
\begin{equation}
a^2+b^2<c^2 \Rightarrow (a+b)^2<c^2+2ab<4c^2 \Rightarrow a+b<2c
\end{equation}
which means $\hat{l}_{s, r}+\hat{l}_{r, t} < 2\hat{l}_{s,t}$.

\subsubsection{The value of $\alpha$}
As shown in Fig. \ref{fig:triangle_rst}, we denote the sides of the triangle formed by the three nodes as $a'$, $b'$ and $c'$ (when $\theta$ is acute). Based on the ASS (Angle-Side-Side) Theorem, we have
\begin{equation}
b'=a' \cos\theta \pm \sqrt{a'^2-c'^2 \sin^2\theta}
\end{equation}
Together with $a'+b'=2c'$ (for the boundary that makes the relationship of $a'+b'$ and $2c'$ different) and solve for $a'$, we have
\begin{equation}
a' = c'\frac{2\cos\theta+\sqrt{2}\sqrt{2\cos^2\theta+\cos\theta-1}+2}{2(\cos\theta+1)}=\alpha c'
\end{equation}

\subsubsection{The value of $\beta$}
Since we want to know the boundary of $\xi^{UB}_{syn}(s, t)=4p^{\hat{l}_{s,t}}q^{\hat{l}_{s,t}-1}$ and $\xi^{UB}_{DODAG}(s, t)=4q^{\hat{l}_{s,r}+\hat{l}_{r,t}-1}$ with respect to $p$ and $q$. We solve for $\hat{l}_{s,t}$ by 
\begin{equation}
4p^{\hat{l}_{s,t}}q^{\hat{l}_{s,t}-1}=4q^{\hat{l}_{s,r}+\hat{l}_{r,t}-1}
\end{equation}
Thus, we have
\begin{equation}
\hat{l}_{s,t} = (\hat{l}_{s,r}+\hat{l}_{r,t}) \frac{\log q}{\log p + \log q} = \beta (\hat{l}_{s,r}+\hat{l}_{r,t})
\end{equation}

\section{Simulation results}
\label{section:simu_res}
We conduct network simulations on the two approaches to support our observation of the theoretical performance analysis. We conduct simulation with single-path scenarios using NextworkX \cite{hagberg_networkx_2023}, where we only find one optimal path for the source and destination in each unit time. We did the single-path scenarios since it is much faster to simulate in a large grid topology. A multi-path scenario is also shown at the end. The locations of the source and destination are generated randomly. The root node for DODAG is set at the center of the grid. To simulate the probabilistic behaviors of direct-link entanglement generation, we employ a grid network with edges that have assigned probabilities for availability. Subsequently, we simulate the entanglement swapping process within the probabilistic vertices of the network. Within this simulated grid network, we conduct 50,000 iterations of each protocol for random pairs located at the same distance, aiming to achieve end-to-end entanglement.

\begin{figure*}
  \includegraphics[width=\textwidth]{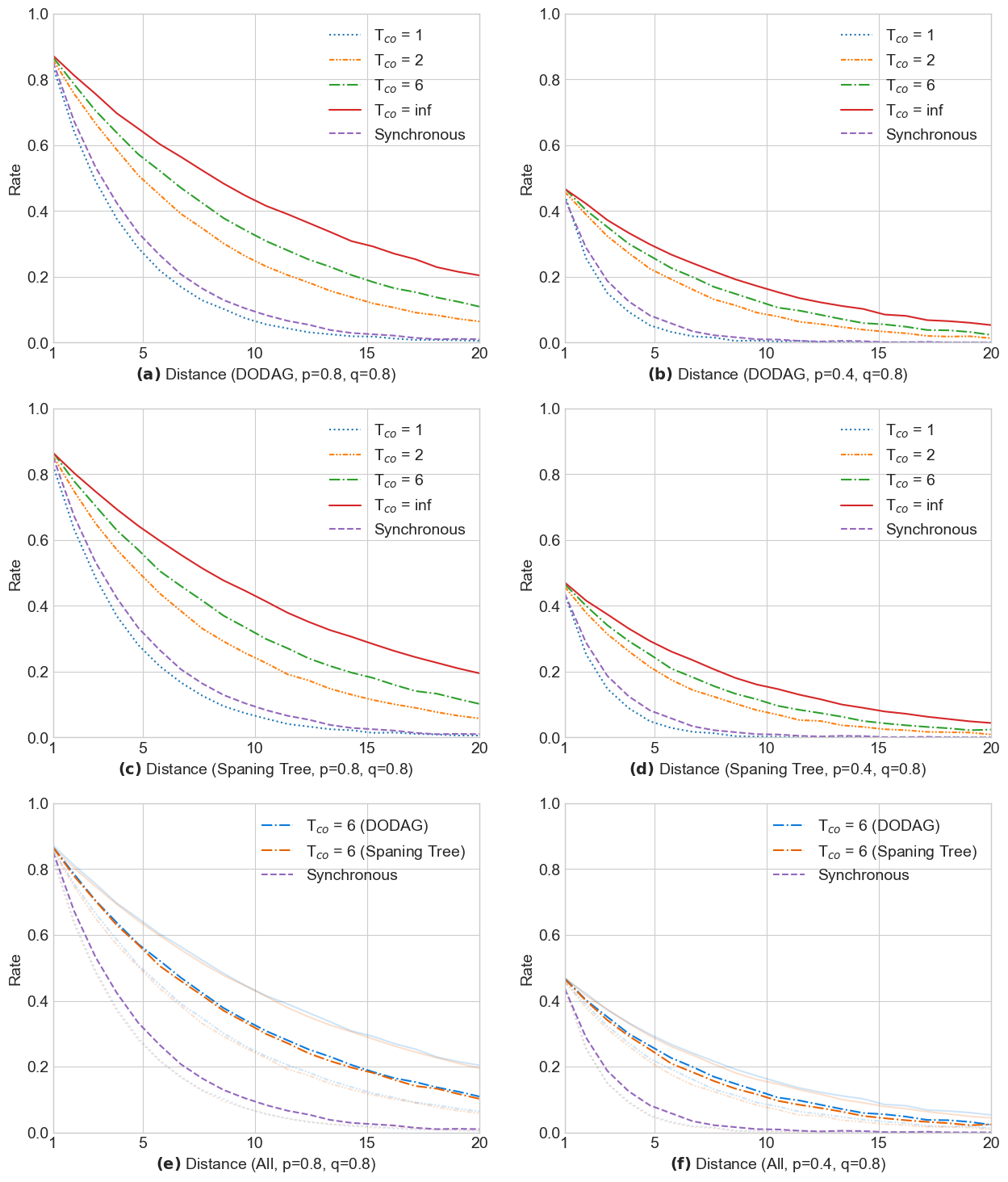}
  \captionsetup{width=1.0\linewidth}
  \caption{\textbf{Simulation results: rate vs. distance.} \textbf{(a)}, \textbf{(c)}, and \textbf{(d)} show the entanglement rate behaviors when $p=q=0.8$, and \textbf{(b)}, \textbf{(d)}, and \textbf{(f)} show those when $p=0.4$ and $q=0.8$ (with local knowledge of the instant topology). The first row, represented by \textbf{(a)} and \textbf{(b)}, showcases the DODAG performance, while the second row, \textbf{(c)} and \textbf{(d)}, highlights the spanning tree performance. These are compared to the performance of synchronous protocols, denoted by the dashed purple lines. The third row, \textbf{(e)} and \textbf{(f)}, includes all protocols together and highlights asynchronous ones when $T_{co}=6$.}
  \label{fig:simu_rd}
\end{figure*}

As shown in Fig. \ref{fig:simu_rd}, the entanglement rates of asynchronous protocols (both DODAG and spanning tree) decay slower than synchronous approaches. We can tell that DODAG and spanning tree approaches behave similarly (DODAG slightly better) in simulation. Both provide higher rates than synchronous approaches when the coherence time is greater than one unit time. We try different combinations of $p$ and $q$ in the simulation. In all combinations, asynchronous protocols have a higher entanglement rate and decay slower than synchronous ones when the coherence time is greater than one unit time. When coherence time is equal to one, asynchronous protocols perform similarly to synchronous protocols when the distance is small but slightly worse when the distance increases. The solid lines in the figure show the estimated upper bounds of the entanglement rates of an asynchronous protocol under the particular parameter settings, which is obtained by using infinite coherence time. The figure also shows that asynchronous protocol has a higher rate even in short-distance scenarios. This verifies our speculation in Section \ref{section:dodagana} that close-by node pairs have a high chance of not needing to go through the root but only through a common ancestor. Note that the simulation results represent the average rate across multiple iterations, which introduces some inherent noise. Additionally, intermediate entanglements (i.e., the intermediate entanglements established towards end-to-end entanglements) set up through entanglement swapping will refresh the coherence time and are retained in asynchronous routing processes. Consequently, the rates in simulations at distance $d$ may exceed $q^d$ when using asynchronous protocols.

\begin{figure*}
  \includegraphics[width=\textwidth]{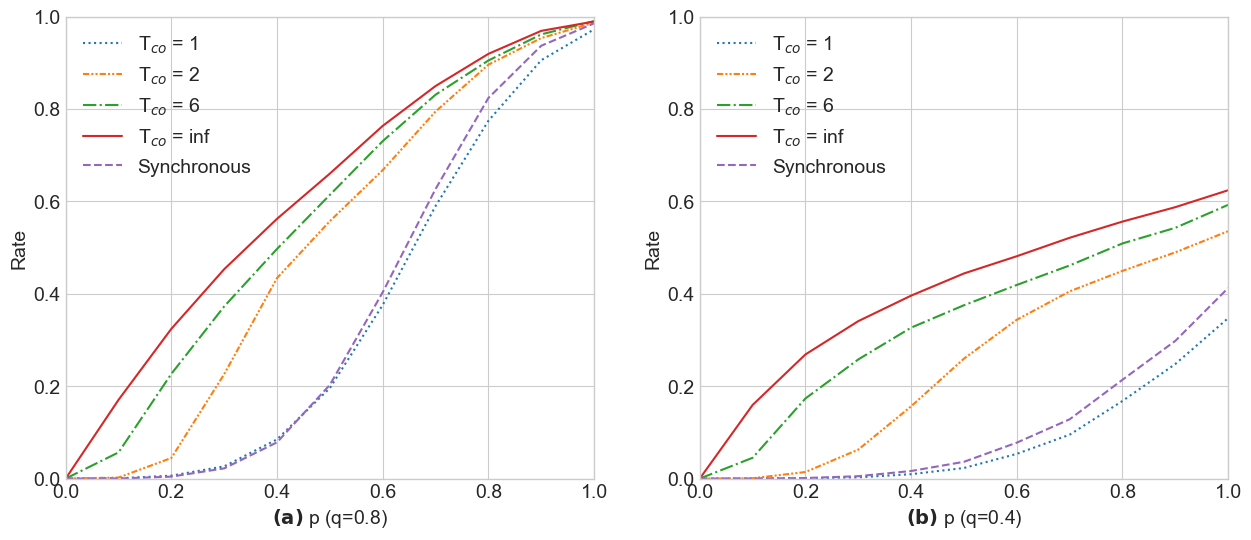}
  \captionsetup{width=1.0\linewidth}
  \caption{\textbf{Simulation results: rate vs. p.} The two diagrams show the relationship between entanglement rate and $p$ in the two types of protocols. For $q=0.8$, this relationship is depicted in \textbf{(a)}, and for $q=0.4$, it is presented in \textbf{(b)}. The coherence time $T_{co}$ is varied across the set $\{1, 2, 6, \infty\}$. When the coherence time is greater than one, the asynchronous protocol has higher rates in all values of $p$. When the coherence time equals one unit time, the asynchronous protocol performs similarly to the synchronous protocol, and its entanglement rate is slightly smaller as $p$ increases. In both situations, the entanglement rate of asynchronous routing increases as coherence time increases, but the synchronous protocol does not.}
  \label{fig:simu_rp}
\end{figure*}

As shown in Fig. \ref{fig:simu_rp}, the entanglement rate of asynchronous routing increases as coherence time increases. The performance of existing synchronous protocols such as \cite{pant_routing_2019} is independent of the coherence time. This indicates that when technology improves, asynchronous protocols will have a much more impact than synchronous protocols. Since the performance of DODAG and spanning tree protocols is similar in simulation, we only show the DODAG graph in this figure for concise. The plot also demonstrates the estimated upper bound of the asynchronous protocol under a particular $q$ when the coherence time is infinite. For example, the upper bounds of the entanglement rate of an asynchronous protocol when $q=0.8$ or $q=0.4$ are shown with the solid line in Fig. \ref{fig:simu_rp}. Nonetheless, we can see if we have $p=q=1$, we do not need any asynchronous or synchronous protocols running in the case of having local knowledge of the instant topology. With that, we can run any shortest path algorithm using the physical topology. However, asynchronous protocols will still be useful, considering a quantum mobile network with dynamic physical topology (i.e., nodes may be moving). Moreover, with $T_{co}=1$, the curves of asynchronous and synchronous protocols look just like the percolation probability in percolation theory and get increasingly obvious (more and more rapid transition around the threshold) when $q$ gets larger.

In this article, we equate quantum state lifetime with the coherence time of entanglement, considering them synonymous due to the dependency of entanglement coherence time on qubit lifetime. As entanglement is sustained by quantum states, its coherence period is determined by how long the states remain coherent. Additionally, we posit that the usable lifetime of qubit—its duration of reusability for entanglement generation—is sufficiently long. Utilizing unit time, which conveys a relative rather than absolute value, prevents us from specifying a minimum quantum state lifetime, prompting us to assume it to be suitably long.

\begin{figure*}
  \includegraphics[width=\textwidth]{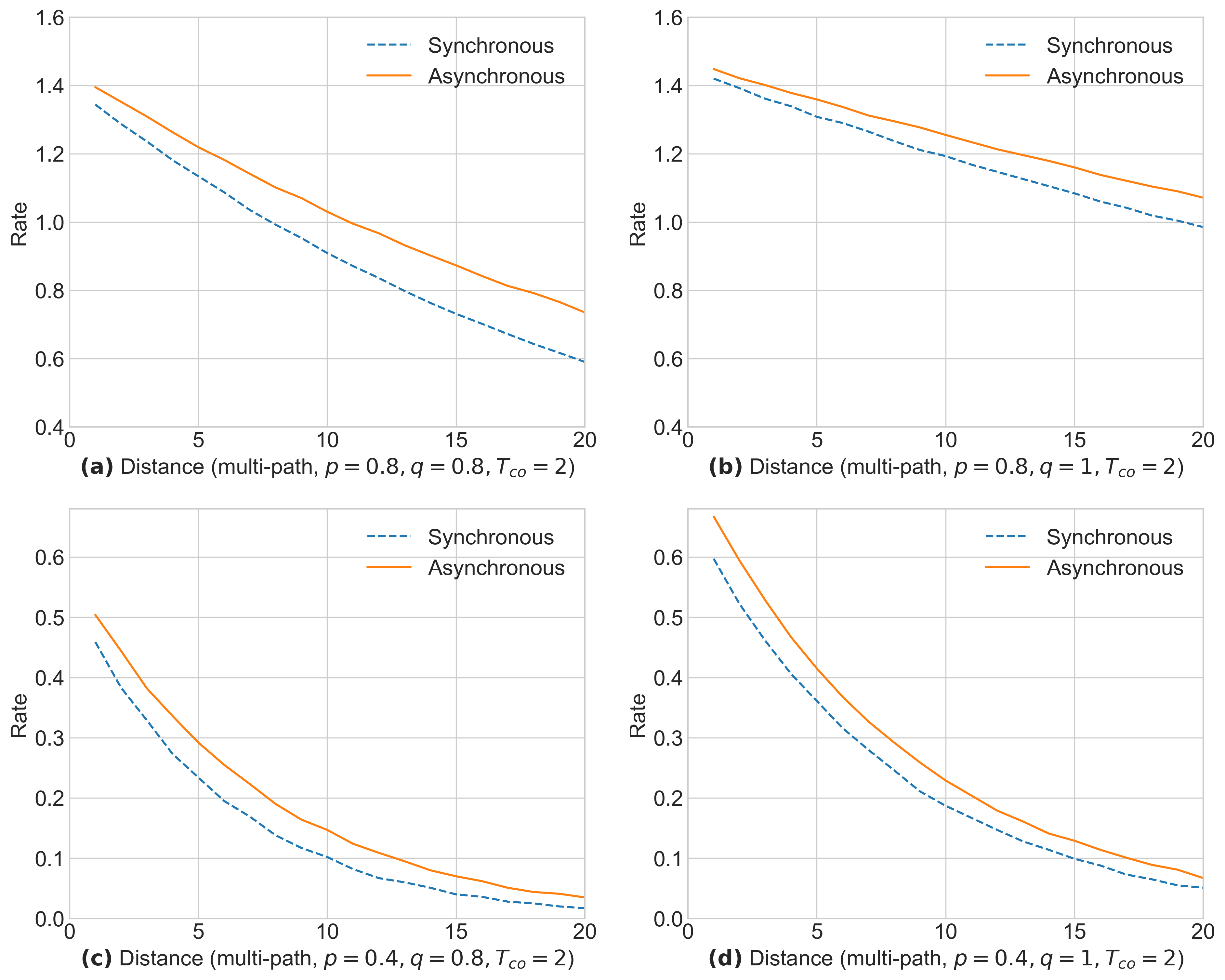}
  \captionsetup{width=1.0\linewidth}
  \caption{\textbf{Multi-path simulation results: rate vs. p.} Previous diagrams demonstrate the performance of single-path scenarios. This figure shows the performance comparisons of multi-path scenarios of the protocols. In varying scenarios with different $p$, $q$, and $T_{co}$ values, asynchronous protocols consistently outperform synchronous ones in terms of entanglement rate. Specifically, for $T_{co}=2$ and $p=q=0.8$, the result is shown in \textbf{(a)}. The scenarios "$p=0.8, q=1$," "$p=0.4, q=0.8$," and "$p=0.4, q=1$" are illustrated in \textbf{(b)}, \textbf{(c)}, and \textbf{(d)} respectively. Overall, our protocol provides larger rates than synchronous approaches in all scenarios when $T_{co}>1$. Note that when $q=1$ and $q>0.5$ (the bond percolation threshold for the square lattice \cite{broadbent_percolation_1957}), the distance affects less on the rate. The rate is not distance independent (it actually is when considering only one path) because there are fewer alternative paths as distance increases. This behavior conforms to what is expected in percolation theory.}
  \label{fig:simu_mp}
\end{figure*}

Finally, we show the simulation results in multi-path scenarios (Fig. \ref{fig:simu_mp}). Multi-path simulation means we search for more than one path (disjoint) for a pair of source and destination. We can tell asynchronous protocols have a higher entanglement rate than synchronous protocols in all scenarios when $T_{co} > 1$.

With the simulation results, we can establish the contributions of our asynchronous protocols:
\begin{enumerate}
  \item The entanglement rate of asynchronous routing increases with the coherence time.
  \item The entanglement rate of asynchronous routing decays slower than synchronous approaches.
  \item The entanglement rate of asynchronous routing is larger than synchronous approaches in almost all scenarios.
\end{enumerate}

\section{Conclusion}
We envision quantum-native routing (i.e., routing for end-to-end entanglement) as the key component of the Quantum Internet and identify the disadvantages of synchronization used in existing quantum-native routing schemes. Thus, we proposed the design of the asynchronous routing scheme and benchmarked its performance under different parameter settings. Here, we highlight the differences between our asynchronous routing protocols and existing asynchronous protocols. Under the condition that a node only knows the direct-link entanglement states of its adjacent links, existing approaches use synchronized time slots, including two execution phases to collaborate all nodes to search for a path connecting the source and destination nodes. This method consumes all direct-link entanglements, trying to find multiple paths between them. However, the consumption of all direct-link entanglements counters the entanglement rate. Our method removes such synchronization of execution phases by introducing graph maintenance in a distributed manner. We treat all network nodes as vertices in a graph that try to connect their neighbors using direct-link entanglements. No central control and no global knowledge of the direct-link entanglements are needed in our method. Whenever a pair of nodes wants to create an end-to-end entanglement, they go through the distributed graph that is continuously updated to find each other. With analysis and simulation, we found that the entanglement rate of our asynchronous protocols increases with the coherence time, decays slower than and is generally larger than that of asynchronous approaches. Furthermore, asynchronous protocols can simultaneously handle multiple user pairs, which remains an open problem in synchronous schemes \cite{pant_routing_2019}. Exploring the performance of multi-user scenarios holds significant prospects for future research.

Moreover, existing approaches treat repeater nodes equally. However, repeaters have different performances (they can have weights for edges, e.g., transmissivity, for routing, but no preference for using competent nodes for more important purposes). Even though this is not the primary purpose, DODAG can reflect the node robustness by preferring to set robust nodes as root nodes. Also, DODAG roots can serve as well-fit gateways to connect multiple networks (i.e., networks of networks). Since DODAG has such benefits that a spanning tree does not have and has similar performance, we recommend DODAG to be the primary protocol for asynchronous routing, which is also well-studied and standardized for low-cost wireless networks. Additionally, the scalability of the DODAG network can be enhanced by constructing hierarchical DODAGs, which involve creating trees of trees. This direction holds potential for future exploration. 

There are other aspects of the protocol that warrant further investigation. For example, the center of a network is a good position for the root node, but the impact of different root positions remains future work. Moreover, the strategies to arrange multiple roots' positions and to determine how fine-grain (the density) the partitioned networks should be (distance between adjacent root nodes) remain future work. Last but not least, other aspects of entanglement-swapping-based routing are also important and under investigation, such as the use of untrusted repeaters, the minimal trust between networks (i.e., no knowledge of the internals of another network), the possible presence of malicious nodes, and the order of entanglement swapping along a path, etc. Furthermore, exploring the performance of asynchronous protocols in non-grid networks is worthwhile. Through this work, we aim to raise awareness and foster advancements in networking techniques for the Quantum Internet.

\section*{Acknowledgments}
This work was supported in part by Cisco University Research Gift \#86944165.

\section*{Data availability}
The data that support the findings of this study are available within the article.

\appendix

\section{Quantum network stack}
\label{appe:networkstack}
\begin{figure}
  \includegraphics[width=\textwidth]{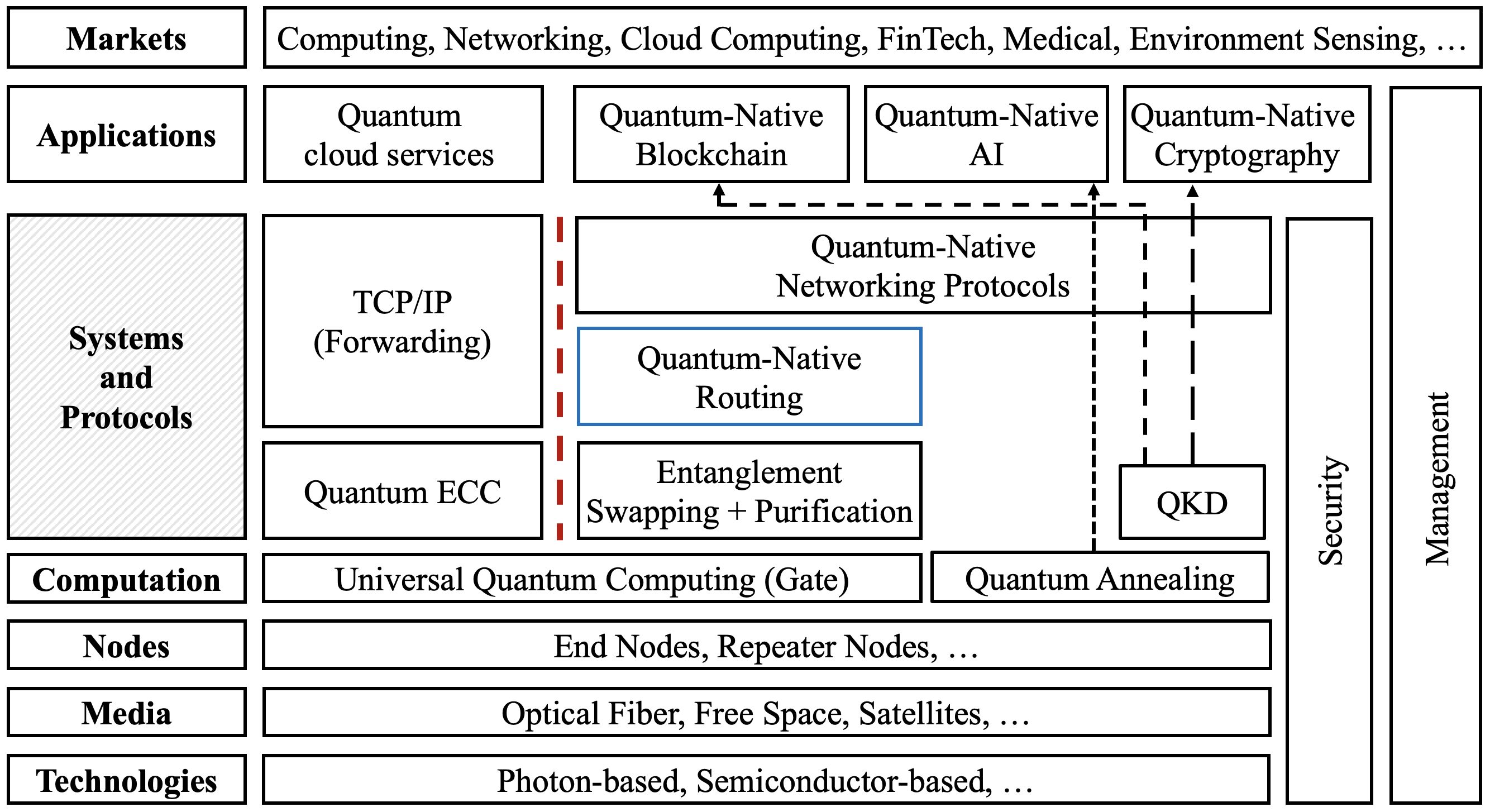}
  \caption{\textbf{7-layer stack of quantum computing research.}}
  \label{fig:qc_stack}
\end{figure}

We have identified seven layers within the technical framework for quantum research, providing a schematic to grasp the diverse elements essential to quantum technology, as illustrated in Fig. \ref{fig:qc_stack}. At the foundational level, the stack is anchored by diverse technologies, which include photon-based and semiconductor-based approaches \cite{yang_survey_2023}, critical for quantum states and operations. These technologies interface with various media, such as optical fiber and free space, such as satellites, providing the physical medium for quantum communication and interactions. Nodes in the quantum network, which encompass end nodes and repeater Nodes, form the fundamental infrastructure, facilitating quantum computation and communication processes. In the computation realm, universal quantum computing (i.e., gate model) and Quantum Annealing are the primary mechanisms that execute quantum algorithms and solve complex problems. Systems and protocols are the backbone of any quantum network. While TCP/IP represents conventional networking standards, the stack introduces quantum-native protocols with unique functionalities like entanglement swapping and asynchronous routing. At the Applications layer, solutions such as quantum cloud services, quantum blockchains, quantum artificial intelligence (AI), and quantum cryptography showcase the potential real-world uses of quantum technologies \cite{lo_secure_2014, yang_pqb_survey_2023, 9893393, 9583587, renduchintala_survey_2022}. Lastly, at the markets layer, the stack emphasizes the wide-ranging sectors that quantum technologies can influence, e.g., computing, networking, clouds, FinTech, medical, and sensing, underlining the vast impact quantum advancements can have across industries.

Our observation shows that while substantial progress has been made in the lower layers, the upper layers have received comparatively less attention. These layers reflect research initiatives in physics and electrical engineering \cite{kimble_quantum_2008, degen_quantum_2017}. In contrast, the upper layers indicate endeavors in software, encompassing network protocols and applications. In particular, within the layer of systems and protocols, there are two distinct approaches: one involves using quantum repeaters that transmit quantum information (left of the red dashed line). The transmission of quantum information through repeaters requires error correction using a significant number of error-correcting qubits, which is restricted by the current capabilities of quantum computers. The other involves distributing entanglement as a communication resource (i.e., quantum-native routing), which is our primary interest and, in our view, deserves greater attention from the research community.

\section{DODAG loop avoidance}
\label{appe:loopavoid}
When nodes select routes in a graph, loops can cause the routing process not to conclude, wasting time and resources. Therefore, they must be detected and avoided. While a directed acyclic graph (DAG) does not form closed loops, a DODAG may do so if a node is greedy and wants to select as many parents as  possible. To prevent this, a rank-based non-greedy mechanism is used to detect loops before allowing a node to join the DODAG.

\begin{figure}
  \includegraphics[width=0.9\textwidth]{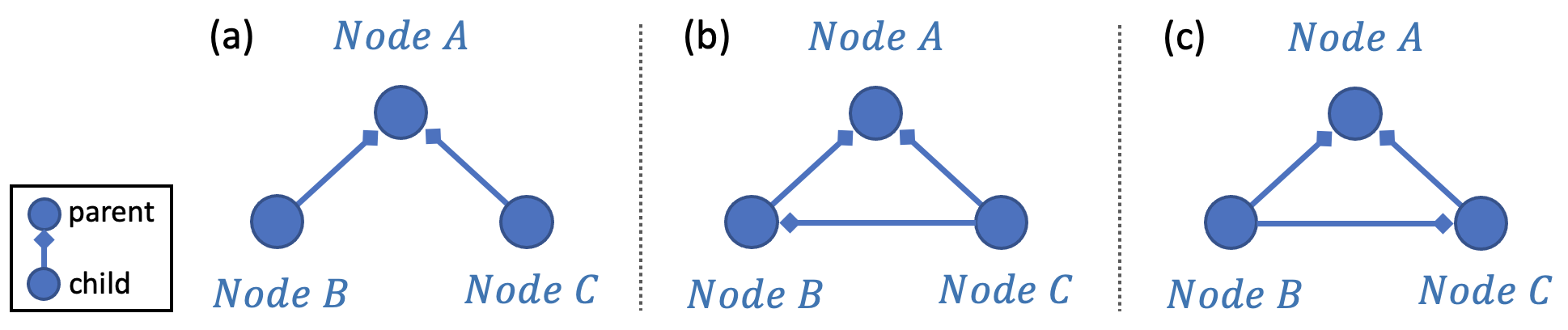}
  \captionsetup{width=1.0\linewidth}
  \caption{\textbf{DODAG greedy parent selection.} \textbf{(a)} The initial state of a branch of a DODAG, where node $A$ has two children, $B$ and $C$. \textbf{(b)} Node $C$ chooses both $A$ and $B$ as parents. \textbf{(c)} Node $B$ intentionally leaves the DODAG and then rejoin at a lower rank, taking both $A$ and $C$ as parents. In this case, \textbf{(b)} and \textbf{(c)} would create an infinite cycle.}
  \label{fig:dodag_loop}
\end{figure}

If a node in the DODAG is greedy and wants to move deeper (closer to the root) by increasing its number of parent nodes, there is a risk of creating a loop. Consider Fig. \ref{fig:dodag_loop}a, which depicts the initial state of a branch of a DODAG, where node $A$ has two children, $B$ and $C$. Let us assume that $A$ is the closest node to the root among the three, making it the preferred parent of $B$ and $C$. If nodes $B$ and $C$ are both eager to select more parents, a loop may occur. For example, node $C$ may want to choose both $A$ and $B$ as parents, making it deeper than both $A$ and $B$, as shown in Fig. \ref{fig:dodag_loop}b. Suppose node $B$ is also greedy and decides to select $C$ as a parent against the non-greedy rule. In this case, node $B$ will intentionally leave the DODAG and then rejoin at a lower rank, taking both $A$ and $C$ as parents, as illustrated in Fig. \ref{fig:dodag_loop}c. As a result, node $B$ will become deeper than both $A$ and $C$. Node $C$, who is also greedy, will then leave and rejoin at a lower rank to get two parents and have a lower rank than both of them. Node $B$ will repeat the process by leaving and rejoining at a lower rank, which will cause the DODAG to alternate between Fig. \ref{fig:dodag_loop}b and c, creating an infinite cycle. This situation can be prevented by ensuring nodes $B$ and $C$ do not select parents from deeper nodes or leave the DODAG intentionally to acquire more parents. Instead, they should maintain a reasonable rank relative to their preferred parent, $A$. Additionally, they should not process parenting requests from deeper nodes. In the example, as long as node $B$ does not depart from the DODAG solely to acquire two parents, the branch of the DODAG will remain in the state depicted in Fig. \ref{fig:dodag_loop}b.

\section{GHS delay scenarios}
\label{appe:ghsdelay}
\begin{figure*}
  \includegraphics[width=\textwidth]{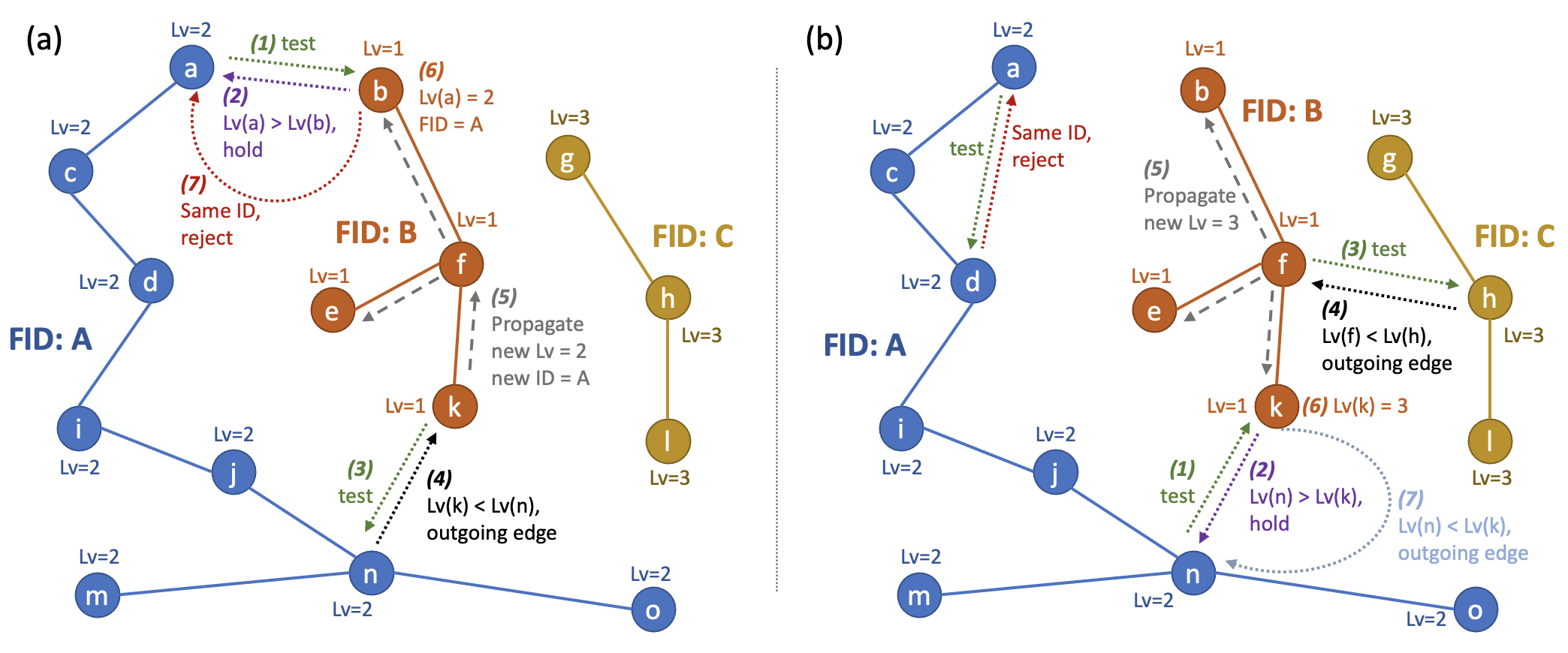}
  \captionsetup{width=1.0\linewidth}
  \caption{\textbf{Fragment combination in the GHS algorithm.} The procedures labeled (1) through (7) operate following the rules outlined in Section \ref{section:dst}. Two scenarios are described as follows: \textbf{(a)} The "test" message from Node $a$ to $b$ is rejected due to the outgoing edge determined at $(4)$. However, if no level value is available to determine whether an edge is outgoing, a loop will be formed through nodes `a-b-f-k-n-j-i-d-c-a.' \textbf{(b)} If the level of the node that initiates a "test" message is less than or equal to the level of the recipient node, then it is safe to consider the edge in-between them outgoing. No loop will be formed in this case. An entanglement link can safely be created for this edge.}
  \label{fig:spanningtree_bc}
\end{figure*}

In Fig. \ref{fig:spanningtree_bc}a, (1) Node $a$ sends a "test" message to Node $b$, but they cannot determine which ID is the latest because $Lv(a)>Lv(b)$. (2) Consequently, they put the edge on hold. (3) Meanwhile, another node, $k$, initiates a "test" message to Node $n$. (4) Since $Lv(k)<Lv(n)$, an outgoing edge is established. (5) Node $k$ propagates the new level and ID to other nodes in the same fragment. (6) Node $b$ now knows that its latest level is $2$ and its latest ID is $A$. (7) Nodes $a$ and $b$ also realize they have the same ID and leave the edge alone. In Fig. \ref{fig:spanningtree_bc}b, a similar process leads to finding an outgoing edge successfully. These processes use level increases to avoid broadcast delay, which may cause loops or unnecessary connections. Thus, fragment splitting while maintaining the level value will not affect the merge or absorb operations in the GHS algorithm. If a link breaks during broadcasting, the on-hold edge will remain on hold until the adjacent nodes broadcast the breaking message. Then, it will release the on-hold status and start the next "test" message. If edge `k-f' breaks in Figures \ref{fig:spanningtree_bc}a and b, only process (2) will remain on hold until the node knows it is isolated. On the other hand, if edge `i-j' breaks, Fragment A splits into two fragments with different IDs, but it will not impact any operations from (1) to (7).

\section{Mean path length approximation}
\label{appe:meanpath}
We can examine the physical 2D grid topology by representing it as a 2D square lattice $G$ comprising of $n$ nodes where $\sqrt{n} \in \mathbb{Z}$. We consider the L1 distance between every two nodes within graph $G$. The L1 distance for nodes $s$ and $t$ with coordinates $s=(x_s, y_s)$ and $t=(x_t, y_t)$ is determined as follows:
\begin{equation}
d(s, t) = d((x_s, y_s), (x_t, y_t)) = |x_s - x_t| + |y_s - y_t|
\end{equation}
The length of the $m$th shortest self-avoiding path between $s$ and $t$ is as follows (c.f., Reference \cite{malik_concurrence_2022}):
\begin{equation}
l^m_{s,t} = d(s, t) + 2(m-1)
\end{equation}
Let us denote the total number of self-avoiding paths of length $l_m$ between $s$ and $t$ as $N^{s,t}_{l_m}$. This value can be obtained by the piecewise path enumeration algorithm introduced in Reference \cite{malik_concurrence_2022}. Then, we can represent the mean length of all the possible paths with maximum length $l_m$ between $s$ and $t$ as
\begin{equation}
l_{s,t} = \frac{\sum^m_{i=1}{l_i N^{s,t}_{l_i}}}{\sum^m_{i=1}{N^{s,t}_{l_i}}}
\end{equation}
Assuming $l_m$ represents the longest path length between nodes $s$ and $t$ in $G$, the above expression corresponds to the average length of all possible paths connecting them. Nevertheless, to simplify the computation without compromising the outcome, we can assign a sufficiently large fixed value to $m$ without necessarily determining the precise average length of all possible paths. Note that the maximum value of $m$ is proportional to $n$.

\bibliographystyle{unsrt}
\bibliography{aipmain}

\end{document}